# Atomic-scale control of graphene magnetism using hydrogen atoms

**Authors:** H. González-Herrero[1], J. M. Gómez-Rodríguez[1,2,3], P. Mallet[4,5], M. Moaied[1,6], J. J. Palacios[1,2,3], C. Salgado[1], M. M. Ugeda[7,8], J.Y. Veuillen[4,5], F. Ynduráin[1,2,3] and I. Brihuega[1,2,3]*.

*Correspondence to: ivan.brihuega@uam.es

[1] Departamento Física de la Materia Condensada, Universidad Autónoma de Madrid, E-28049 Madrid, Spain.

[2] Condensed Matter Physics Center (IFIMAC), Universidad Autónoma de Madrid, E-28049 Madrid, Spain.

[3] Instituto Nicolás Cabrera, Universidad Autónoma de Madrid, E-28049 Madrid, Spain

[4] Université Grenoble Alpes, Institut NEEL, F-38042 Grenoble, France

[5] CNRS, Institut NEEL, F-38042 Grenoble, France.

[6] Department of Physics, Faculty of Science, Zagazig University, 44519 Zagazig, Egypt.

[7] CIC nanoGUNE, 20018 Donostia-San Sebastian, Spain.

[8] Ikerbasque, Basque Foundation for Science, 48011 Bilbao, Spain.

**Isolated hydrogen atoms absorbed on graphene are predicted to induce magnetic moments. Here we demonstrate that the adsorption of a single hydrogen atom on graphene induces a magnetic moment characterized by a ~20 meV spin-split state at the Fermi energy. Our scanning tunneling microscopy (STM) experiments, complemented by first-principles calculations, show that such a spin-polarized state is essentially localized on the carbon sublattice complementary to the one where the H atom is chemisorbed. This atomically modulated spin-texture, which extends several nanometers away from the H atom, drives the direct coupling between the magnetic moments at unusually long distances. Using the STM tip to manipulate H atoms with atomic precision, we demonstrate the possibility to tailor the magnetism of selected graphene regions.**

Adding magnetism to the long list of graphene's capabilities has been pursued since its discovery (*1*). From a theoretical point of view, inducing magnetic moments in graphene rests on removing a single $p_z$ orbital from the $\pi$−graphene system; this removal creates a single $\pi$-state at the Fermi energy ($E_F$) around the missing orbital. The double occupation of this state by two electrons with different spin is forbidden by the electrostatic Coulomb repulsion, namely, once an electron occupies the state, a second one with opposite spin needs to pay an extra energy $U$. This leaves a single electron occupying the state and therefore a net magnetic moment (*2-6*). The strength of $U$, which determines the spin-splitting, depends on the spatial localization of the state since this defines the proximity between the electrons (see Fig. 1A). Contrary to magnetic moments of strong localized atomic character commonly found in magnetic materials, these induced moments are predicted to extend over several nanometers, anticipating a strong direct coupling between them at unusually long distances. The coupling rules between the induced magnetic moments are also expected to be simple. Because of the bipartite atomic structure of graphene, consisting of two equivalent triangular sublattices A and B, and according to Lieb's theorem (*7*), the ground state of the system possesses a total spin given by $S=1/2\cdot|N_A-N_B|$, where $N_A$ and $N_B$ are the number of $p_z$ orbitals removed from each sublattice (*4, 8, 9*). Thus, to generate a net magnetic moment in a particular graphene region, a different number of $p_z$ orbitals from each sublattice needs to be locally removed.

Many theoretical proposals have been put forward in this regard, involving zigzag edges, graphene clusters, grain boundaries, and atomic defects (*2, 4, 5, 8-11*). Experimentally, the removal of $p_z$ orbitals from the $\pi$ system has been achieved by randomly creating atomic vacancies or adsorbing adatoms (*12-16*). However, removing those $p_z$ orbitals in a controlled manner has turned out to be challenging. Here, we rely on the simplest, albeit demanding, experimental approach to remove a single $p_z$ orbital from the graphene network by the adsorption of a single H atom. Atomic H chemisorbs on graphene on top of carbon atoms, changing the initial $sp^2$ hybridization of carbon to an essentially $sp^3$ (*17, 18*) and effectively removing the corresponding $p_z$ orbital (*4, 19, 20*). In this sense, chemisorbed H atoms are equivalent to carbon vacancies (*4, 12, 14*) with the advantage that, contrary to vacancies, they leave the graphene atomic lattice with no unsaturated dangling bonds, preserving the three-fold symmetry. Our experiments, supported by ab-initio calculations, provide a comprehensive picture for the origin, coupling, and manipulation of the magnetism induced by H atoms on graphene layers.

We deposited atomic H on graphene grown on a SiC(000-1) substrate (*21*). In this system the rotational disorder of the graphene layers electronically decouples π bands leading to a stacking of essentially isolated graphene sheets (*22-24*). STM visualizes single H atoms as a bright protrusion (~2.5 Å apparent height) surrounded by a complex threefold $\sqrt{3}\times\sqrt{3}R30°$ pattern (*25, 26*) (see Fig. 1B and Sect. 1 in (*27*)). The resolution achieved here allows us to identify the adsorbate nature and the atomic site -thus the corresponding atomic sublattice- where each H atom is chemisorbed, by comparison with density functional theory (DFT) simulated STM images (see Fig. 1D and Sect. 1 in (*27*)). As sketched in Fig. 1A, graphene magnetic moments induced by H adsorption should be reflected in the appearance of a spin-polarized state at the Fermi energy which, according to DFT calculations, is characterized by two narrow peaks in the density of states ( (*4*) and Fig. 1E).

Differential conductance spectra (*dI/dV*) probe the energy-resolved local density of states (LDOS($E$)) under the tip position and thus is an ideal tool to investigate this central question. Figure 1C shows two *dI/dV* spectra, measured at 5 K, summarizing our findings. *dI/dV* spectra measured on clean graphene, far enough from defects, shows the characteristic featureless "V shape" of graphene with a minimum at $E_F$ indicating the position of the Dirac point $E_D$. *dI/dV* spectra measured on top of single H atoms reveal the existence of two narrow peaks, one below and one above $E_F$, separated in energy by a splitting of ~20 meV. We attribute these two features to the spin-polarized state, in which the Coulomb repulsion is large enough to fully separate the two spin components. The observed charge neutrality (the splitting is essentially symmetric around $E_F$) and the well-defined peak splitting indicates the complete spin-polarization of the state. DFT calculations show that the magnetic moment associated with the unpaired electron left over in the graphene system after H adsorption would be 1μB (see Fig. 7 in (27)) Our interpretation of the experiment is fully supported by DFT calculations, as can be seen in Fig. 1E, which shows the expected DOS for a single H atom in a 218 carbon atoms graphene super cell. The theoretical energy splitting depends on the size of the graphene super cell (*5*). Our calculations shows that the splitting decays with the size of the graphene super cell suggesting a small, but finite, splitting for the isolated H (see Sect. 9 in (*27*)) in agreement with the experimental observations.

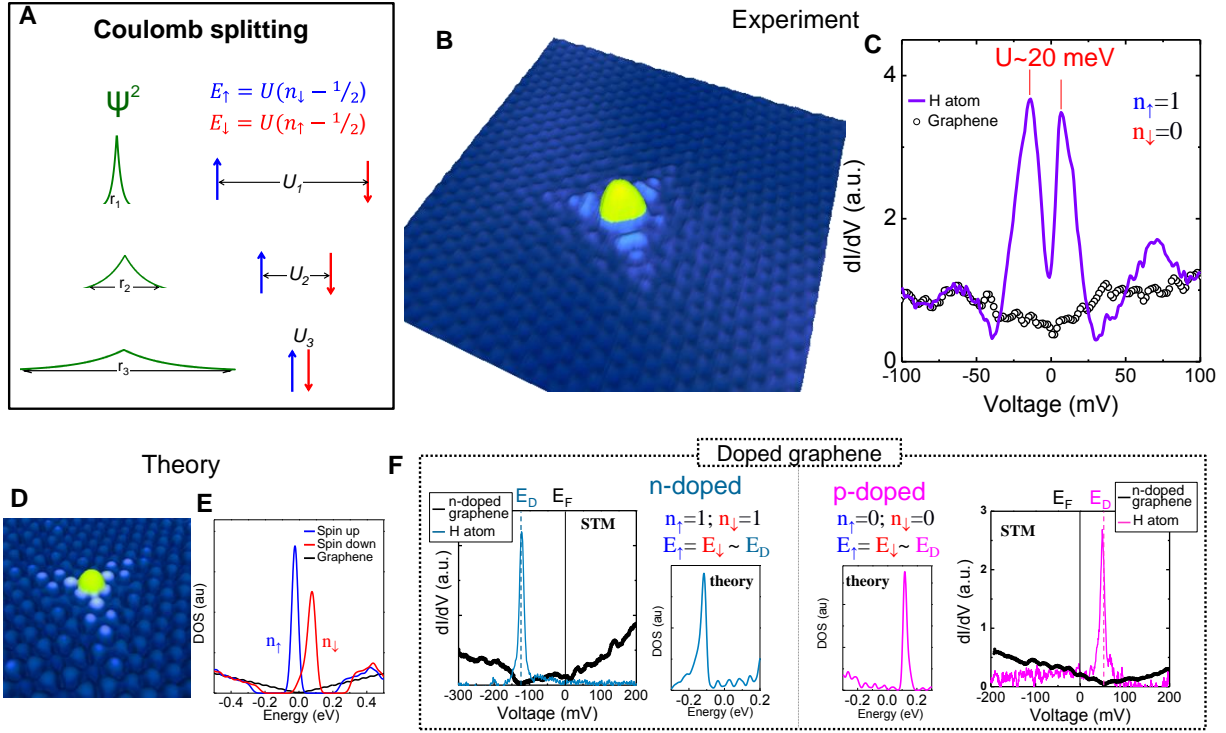

**Fig. 1. Spin split state induced by atomic H on graphene**. (A) Illustration of the origin of the spin-split state in terms of its spatial localization, given by the square of its wave function $|\psi|^2$, and the corresponding electrostatic Coulomb repulsion $U$. Arrows indicate the energy position of spin-up and spin-down levels. For a fully polarized one-electron state the majority level spin is filled and the minority one is empty; therefore $n_\uparrow=1$ and $n_\downarrow=0$, and the energy splitting ($E_\downarrow-E_\uparrow$) is given by $U$ (see Sect. 3 in (*27*)). (B) STM topography of a single H atom chemisorbed on neutral graphene (0.2V, 0.1nA, 7x7nm$^2$). (C) *dI/dV* spectrum on the H atom, showing the appearance of a fully polarized peak at $E_F$, and of bare graphene far from the H atom. The spectra were acquired at a nominal junction impedance of 2 GΩ (-100 mV, 50 pA). (D, E) DFT simulated STM image (D) and DOS (E) of a H atom chemisorbed on neutral graphene. (F) *dI/dV* spectra and DFT calculation of the DOS induced by a single H atom on n- (left panel) and p-type (right panel) doped graphene. The minimum of the *dI/dV* spectra acquired on bare graphene (black curves) determines the position of $E_D$. The spectra were acquired at a nominal junction impedance of 8 GΩ (-400 mV, 50 pA). STM data were acquired and processed using the WSxM software (*36*). All simulated images are calculated at the same energy as the corresponding experimental one. All experimental data were acquired at 5 K.

An independent proof of the magnetic nature of the *dI/dV* split-peaks can be obtained by changing the occupation of the split states ($n_\downarrow$, $n_\uparrow$) via graphene-doping. This follows the idea originally proposed in Ref. (*28*), according to which the transition from a magnetic state to a non-magnetic one can be realized by tuning the energy position of the impurity level with respect to the Fermi level (see Sect. 3a in (*27*) for a detailed description of this system in terms of the Anderson impurity model). In graphene, the impurity level (zero energy mode) should follow the position of $E_D$ (*29*) [$E_\uparrow=E_D+U(n_\downarrow - 1/2)$; $E_\downarrow =E_D+U(n_\uparrow - ½)$ ] which can be tuned by doping the graphene layers. For large enough electron (hole) doping, the occupation of both $n_\uparrow$ and $n_\downarrow$ levels can be tuned to 1 (0), in which case the energy levels will be degenerate, leading to a single non-magnetic state close to $E_D$ ($E_\uparrow = E_\downarrow \approx E_D$). In Fig. 1F we show how graphene-doping affects the splitting of the H induced magnetic state. Our *dI/dV* spectra show that both n- and p-type graphene-doping cause the splitting of the H induced graphene state to vanish; only one sharp peak appears at $E_D$, which we ascribe to a transition to the non-magnetic state. This interpretation is fully supported by our DFT calculations for doped graphene layers (Fig. 1F and Sect. 3b in (*27*)). If the split peaks appearing in neutral graphene after H adsorption had an origin not associated with a magnetic moment (i.e. a single electron origin) we would observe a rigid shift of the peak position with doping, the doublet structure remaining unmodified (see Sect. 3 in (*27*)). Our results are consistent with the case of sp$^3$ defects in graphene, where the possibility to control graphene magnetic moments by molecular doping was reported (*15*).

We have explored, with atomic precision, the spatial extension of the spin-polarized electronic state induced by H atoms in undoped graphene. The relatively modest 20 meV energy splitting observed in our experiments suggests a large spatial extension of the magnetic state (Fig. 1A). Figure 2A shows a conductance map vs energy, *dI/dV(x, E)*, measured along the 6 nm line across the H atom drawn in Fig. 2B. The state extends several nanometers away from the H atom indicating that it is indeed a quasilocalized graphene state (*3*). It presents strong atomic-scale modulations of the peak intensities, with maxima (bright yellow features in the *dI/dV* map) corresponding to C atoms of the opposite sublattice with respect to H chemisorption. On C-sites of the same sublattice, the peaks vanish almost completely (see Fig. S11 in (*27*) for more details). Because our results show a complete spin-polarization of the state, the spatial evolution of the *dI/dV* occupied peak height provides the spatial distribution of the local magnetic moment induced by H chemisorption (see DFT calculations in Figs. 2, C and D and Sect. 4 in (*27*)). This is further plotted in Fig. 2E, which shows that the magnetic moment is essentially induced on the carbon atoms belonging to the graphene sublattice opposite to H adsorption.

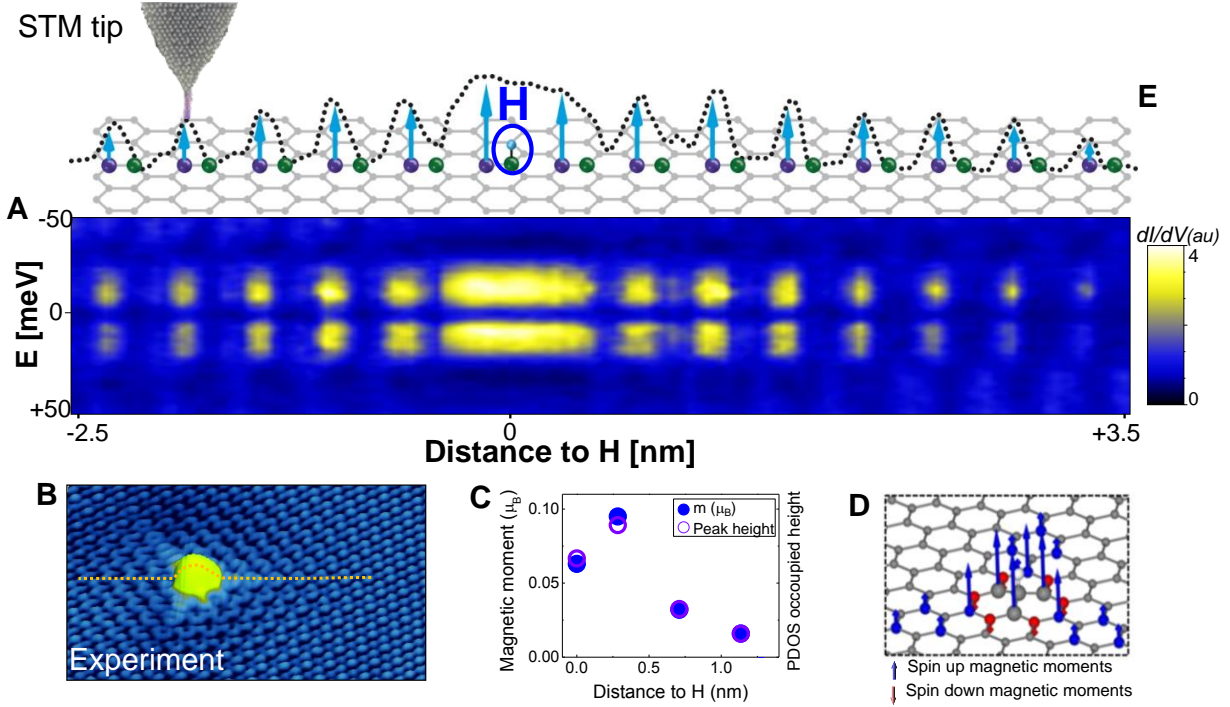

**Fig2. Spatial extension of the spin-polarized electronic state induced by H atoms in undoped graphene** (A) Conductance map -$dI/dV(x, E)$- along the dashed line drawn in (B) The spectra were acquired at a nominal junction impedance of 3 GΩ (100mV, 33 pA). (B) STM topography of a single H atom on graphene (0.2V, 0.1nA, 7x5nm$^2$). (C) Comparison between DFT calculations for the local magnetic moment and occupied PDOS peak height calculated on different C atoms, see Sect. 4 in (*27*). (D) Calculated magnetic moments induced by H chemisorption. (E) Schematics of the graphene structure along the profile drawn in (B), green/purple balls indicating the positions of C atoms belonging to the same/opposite sublattice with respect to the site of the H chemisorption. The dotted line shows the evolution of the measured occupied peak height and the arrows the relative magnetic moment contribution of each C atom (see text). All experimental data were acquired at 5 K.

We now focus on the interactions between the magnetic moments induced in graphene by neighboring H atoms. The large extension of the local magnetic moments associated with H chemisorption suggests that long-range magnetic interactions mediated by direct exchange should take place. This is different from substrate-mediated interactions such as RKKY because here the coupling results from direct overlap of magnetized graphene states. In addition, the critical C sublattice dependence observed for the spin-polarized peak implies that the magnetic coupling should be radically different between H atoms adsorbed on the same and different sublattices. Consistent with this expectation, our DFT calculations reveal that two H atoms chemisorbed on the same sublattice (AA dimer) show ferromagnetic coupling with a total spin S=1, whereas for H atoms on different sublattices (AB dimer) the solution is non-magnetic. This result is reproduced for all possible H-H arrangements up to the largest distances (of ~1.5nm) achievable with our supercell sizes, see Fig. 3A. For a separation of 1.5 nm, the computed exchange energy for AA dimers is ~35 meV (Sect. 5 in (*27*)). Furthermore, the total energy of all H dimers studied is found to be lower than that of two isolated H atoms (Fig. 3A), confirming the observed tendency of H to form dimers on graphene surfaces for high enough H concentrations (*30-32*).

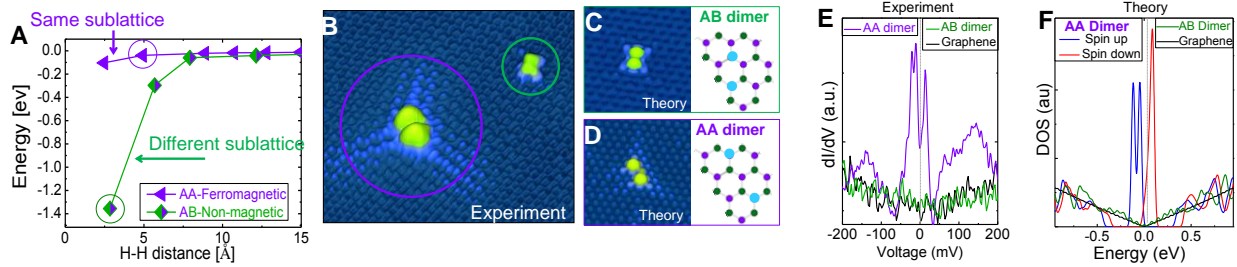

**Fig. 3. Sublattice dependence of the magnetic coupling between neighboring H atoms**. (A) Calculated total energy, relative to twice the adsorption energy of a single H atom, and magnetic state of a pair of H atoms adsorbed on the same (AA) and different (AB) sublattices, plotted as a function of the H-H distance. (B) STM image showing two different pairs of H atoms in an AA (purple circle) and AB (green circle) configuration (0.2V, 0.1nA, 7.8x6.6nm$^2$). (C, D) Calculated STM images of the AB and AA dimer outlined in (A) and shown in (B) and the corresponding diagrams for H atoms (blue balls) on graphene (purple and green balls). (E, F) Experimental *dI/dV* spectra (E) and calculated DOS (F) for the AA dimer, AB dimer and clean graphene. The spectra were acquired at a nominal junction impedance of 8 GΩ (-400 mV, 50 pA). All experimental data were acquired at 5 K.

To test this scenario experimentally, we have explored the local electronic structure of many different H dimers with high-resolution scanning tunneling spectroscopy (STS) measurements. The STM image in Fig. 3B shows two H dimers in AA (purple circle) and AB (green circle) configurations; the corresponding calculated STM images are in Figs. 3, C and D. *dI/dV* spectra acquired on the AB dimer (green line in Fig. 3E) show a featureless LDOS indistinguishable from that taken on bare graphene (black), in good agreement with a non-magnetic configuration. In contrast, *dI/dV* spectra measured on the AA pair (purple), show the split-state in the vicinity of $E_F$, as expected for a ferromagnetic coupling between the H atoms. As shown in Fig. 3F, our calculated DOS reproduce these observations confirming the ferromagnetic (non-magnetic) nature of the AA (AB) dimer. Again, this is a very robust result, and the *dI/dV* spectra of all H dimers measured here showed this behavior: AA dimers present a fully split-state close to $E_F$, which is absent in AB dimers, see Fig. S12 in (*27*). Our STS data show that this sublattice dependent magnetic coupling persists for very long distances, even for H dimers separated by more than 1nm (Sect. 5 in (*27*)).

We now demonstrate the capability of inducing magnetic moments on selected graphene regions by using STM to perform atomic manipulations (*33-35*). We prove that individual H atoms can be removed, laterally moved, and even deposited on graphene surfaces with atomic precision to ultimately tailor their local magnetic state (see Sect. 6 in (*27*) for details). Figure 4 shows two representative examples of such manipulation experiments where the local graphene magnetism was selectively switched ON and OFF. The graphene region in Fig. 4A shows two H atoms in an AB dimer configuration. Our STS data taken on those H atoms (Fig. 4C) show that such an AB dimer does not induce any magnetism on the graphene layer, in good agreement with the coupling rules previously discussed. Figure 4B shows the same graphene region as in Fig. 4A after the controlled extraction of one H atom by gently approaching the STM tip towards it. As shown in Fig. 4D, a spin-split state immediately emerges on the graphene layer after the H removal, confirming the creation of a local magnetic moment in graphene. The insets show the corresponding DFT calculations of the resulting magnetic moment for each situation. Next, we present a lateral manipulation performed on the H dimer shown in the central region of Fig. 4E. Initially, the dimer was in an AA configuration with both H atoms chemisorbed on the same carbon sublattice. The STS spectrum for that configuration (Fig. 4G) shows the presence of a spin-split state, as expected from ferromagnetic coupling. In order to switch off the graphene magnetic moments induced by this H dimer, we turned it into a non-magnetic AB dimer by laterally moving one of its H atoms from one sublattice to the opposite. Figure 4F corresponds to exactly the same graphene region after the H manipulation (the AB dimer in the upper part of the image serves as reference). Our STS spectrum measured on the constructed AB dimer shows the disappearance of the polarized peaks indicating that local graphene magnetism was effectively switched off.

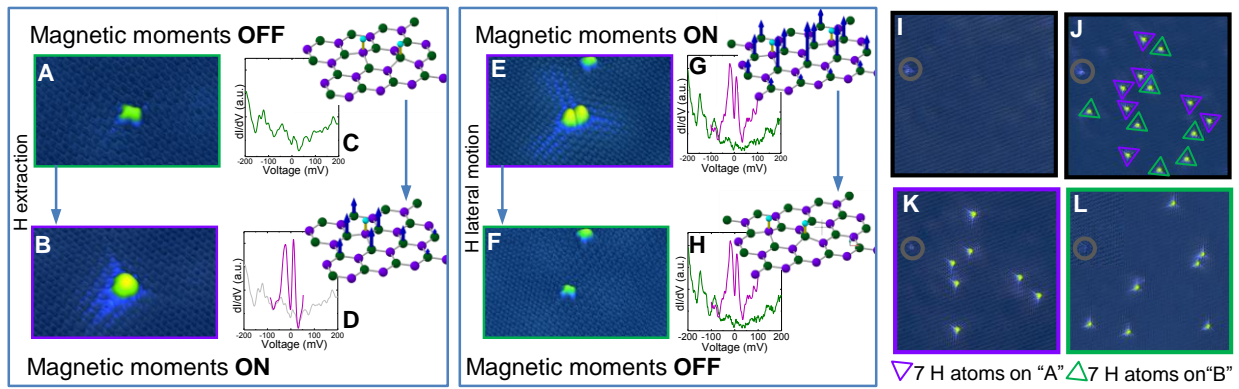

**Fig. 4. Manipulation of graphene local magnetic moments by STM**. (A) STM image of an H dimer in an AB configuration. (B) STM image after the removal of one H atom. Insets present the corresponding DFT calculation for H atoms (blue balls) on graphene (purple and green balls), with blue arrows being the magnetic moments induced on graphene. (C, D) *dI/dV spectra* measured on the AB dimer in (A) and the single H atom in (B) respectively. The spectra were acquired at a nominal junction impedance of 4 GΩ (200 mV, 50 pA). (E) STM image of a H dimer in an AA configuration. (F) STM image after laterally moving one H atom. Insets are DFT calculations. (G, H) *dI/dV spectra* measured on the AA dimer in (E) and the AB dimer in (F) respectively. The spectra were acquired at a nominal junction impedance of 4 GΩ (200 mV, 50 pA). An additional *dI/dV* spectrum -better resolved in the vicinity of $E_F$- measured on the AA dimer in (E) is shown in Fig. S20 in (*27*). (I-L) STM images showing exactly the same graphene region for different steps of a manipulation experiment involving a large number of H atoms (see text). The point defect outlined with a grey circle is used as a reference. Tunneling parameters: (0.2V, 0.1nA, 6.5x4.0nm$^2$) for (A,B), (0.2V, 0.1nA, 9.5x5.5nm$^2$) for (E, F), (0.4V, 0.03nA, 28x28nm$^2$) for (I-L). All experimental data were acquired at 5 K.

Finally, we explore the possibility of selectively tuning the collective magnetic moment in a graphene region by inducing an imbalance between H atoms on A and B sublattices. For this purpose, we systematically manipulated a large number of H atoms (see Sect. 5 in (*27*) for more details). In Figs. 4I-L we present an example where we first removed all H atoms from a graphene region using the STM tip (Fig. 4I). Then, we selectively deposited 14 H atoms on this same region to reach a configuration with 7 H atoms chemisorbed on each graphene sublattice (Fig. 4J). Our experimental findings and existing calculations (*4, 7*) indicate that a very low, if any, net magnetic moment is expected on this region because of this equal sublattice population. Next, by selectively removing all the H atoms chemisorbed on one particular sublattice B we end up in a ferromagnetic configuration with the 7 remaining H atoms on the same sublattice A (Fig. 4K). As a final step, we combined several manipulation processes to reverse the situation and construct an H arrangement with all the 7 H atoms chemisorbed on the opposite sublattice B (Fig. 4L). The degree of complexity shown in our manipulation experiments demonstrates the high reproducibility of the procedure, which paves the way to the realization of atomically controlled graphene magnetism experiments restricted so far to a pure theoretical framework.

**References and Notes:**

1. K. S. Novoselov *et al.*, *Science* **306**, 666 (2004).
2. P. O. Lehtinen, A. S. Foster, Y. C. Ma, A. V. Krasheninnikov, R. M. Nieminen, *Physical Review Letters* **93**, 187202 (2004).
3. V. M. Pereira, F. Guinea, J. dos Santos, N. M. R. Peres, A. H. C. Neto, *Physical Review Letters* **96**, 036801 (2006).
4. O. V. Yazyev, L. Helm, *Physical Review B* **75**, 125408 (2007).
5. O. V. Yazyev, *Reports on Progress in Physics* **73**, 056501 (2010).
6. E. J. G. Santos, A. Ayuela, D. Sanchez-Portal, *New Journal of Physics* **14**, 043022 (2012).
7. E. H. Lieb, *Physical Review Letters* **62**, 1201 (1989).
8. J. Fernandez-Rossier, J. J. Palacios, *Physical Review Letters* **99**, 177204 (2007).
9. J. J. Palacios, J. Fernández-Rossier, L. Brey, *Physical Review B* **77**, 195428 (2008).
10. M. Fujita, K. Wakabayashi, K. Nakada, K. Kusakabe, *Journal of the Physical Society of Japan* **65**, 1920 (1996).
11. C. G. Tao *et al.*, *Nature Physics* **7**, 616 (2011).
12. M. M. Ugeda, I. Brihuega, F. Guinea, J. M. Gomez-Rodriguez, *Physical Review Letters* **104**, 096804 (2010).
13. R. R. Nair *et al.*, *Nature Physics* **8**, 199 (2012).
14. K. M. McCreary, A. G. Swartz, W. Han, J. Fabian, R. K. Kawakami, *Physical Review Letters* **109**, 186604 (2012).
15. R. R. Nair *et al.*, *Nature Communications* **4**, 2010 (2013).
16. J. Balakrishnan, G. K. W. Koon, M. Jaiswal, A. H. C. Neto, B. Ozyilmaz, *Nature Physics* **9**, 284 (2013).
17. L. Jeloaica, V. Sidis, *Chemical Physics Letters* **300**, 157 (1999).
18. X. W. Sha, B. Jackson, *Surface Science* **496**, 318 (2002).
19. D. W. Boukhvalov, M. I. Katsnelson, A. I. Lichtenstein, *Physical Review B* **77**, 035427 (2008).
20. D. Soriano *et al.*, *Physical Review Letters* **107**, 016602 (2011).
21. F. Varchon, P. Mallet, L. Magaud, J. Y. Veuillen, *Physical Review B* **77**, 165415 (2008).
22. M. Sprinkle *et al.*, *Physical Review Letters* **103**, 226803 (2009).
23. G. H. Li *et al.*, *Nature Physics* **6**, 109 (2010).
24. G. T. de Laissardiere, D. Mayou, L. Magaud, *Nano Letters* **10**, 804 (2010).
25. H. A. Mizes, J. S. Foster, *Science* **244**, 559 (1989).
26. P. Ruffieux, O. Groning, P. Schwaller, L. Schlapbach, P. Groning, *Physical Review Letters* **84**, 4910 (2000).

27. Materials and methods are available as supporting material on *Science* Online
28. P. W. Anderson, *Physical Review* **124**, 41 (1961).
29. B. Uchoa, V. N. Kotov, N. M. R. Peres, A. H. C. Neto, *Physical Review Letters* **101**, 026805 (2008).
30. L. Hornekaer *et al.*, *Physical Review Letters* **97**, 186102 (2006).
31. Z. Sljivancanin *et al.*, *Journal of Chemical Physics* **131**, 084706 (2009).
32. R. Balog *et al.*, *Journal of the American Chemical Society* **131**, 8744 (2009).
33. D. M. Eigler, E. K. Schweizer, *Nature* **344**, 524 (1990).
34. T. C. Shen *et al.*, *Science* **268**, 1590 (1995).
35. A. A. Khajetoorians *et al.*, *Science* **339**, 55 (2013).
36. I. Horcas *et al.*, *Review of Scientific Instruments* **78**, 013705 (2007).

**Acknowledgments:** We thank V. Cherkez, from Institut Néel, CNRS-UJF, for his help during the graphene on SiC samples fabrication and Dillon Wong for his careful reading of the manuscript. This work was supported by Spain's MINECO under Grants No. MAT2013-41636-P, No. CSD2010-00024, PCIN-2015-030; FIS2013-47328 and FIS2012-37549-C05-03; the European Union structural funds and the Comunidad de Madrid MAD2D-CM Program (S2013/MIT-3007) and P2013/MIT-2850: the Generalitat Valenciana under Grant No. PROMETEO/2012/011; the CNRS DERCI PICS n°6182 and the European Union FP7 "Graphene Flagship" program (Grant No. 604391) and FLAG-ERA program. The authors acknowledge the computer resources and assistance provided by the Centro de Computación Científica of the Universidad Autónoma de Madrid.

**Supplementary Materials:**

Materials and Methods

Figs. S1 to S20

Captions for Movies S1

References (37-59)

# Supplementary Materials for

## Atomic-Scale Control of Graphene Magnetism Using Hydrogen Atoms."

Héctor González-Herrero, José M. Gómez-Rodríguez, Pierre Mallet, Mohammed Moaied, Juan J. Palacios, Carlos Salgado, Miguel M. Ugeda, Jean-Yves Veuillen, Félix Ynduráin, Iván Brihuega[*].

*correspondence to: ivan.brihuega@uam.es

**This PDF file includes:**

    Materials and Methods
    Figs. S1 to S20
    Captions for Movies S1
    Full Reference List

**Other Suplementary Materials for this manuscript includes the following:**

    Movie S1

**Materials and Methods**

**1. Sample preparation and experimental details.**

A key point of the present work is the atomistic control of the samples, which was obtained by performing all the preparation procedures and measurements under UHV conditions (during the whole process -imaging pristine graphene sample => depositing H atoms on it => and imaging it back- the sample was maintain in the same UHV system).

The samples are prepared under UHV by graphitization of a 6H-SiC(000-1) surface (*21, 37*). The substrate surface is cleaned under a Si flux and subsequently annealed at 950-1000°C to get the 6H-SiC(000-1) (3x3) surface reconstruction. Further annealing at higher temperature induces the growth of (rotationnaly disordered) graphene layers on this surface (*21*). The onset of graphitization is detected by LEED (*37*). The average number of graphene layers on the surface is governed by the annealing temperature/time and is controlled by Auger electron spectroscopy (*21*). A thick (more than 5 graphene layers) and thus surface neutral multilayer was grown for the studies reported in the body of the paper. A thinner one (average thickness of 2-3 graphene layers) whose surface layer is doped by interfacial charge transfer (see Sect. 3c) was prepared for studying the influence of the n-doping on the H induced state. Graphene p-doping was achieved after exposing our samples to many cycles of H dosing and further annealing (see Sect. 3c).

We deposited atomic hydrogen following the procedure of refs (*30-32*), i.e. by the thermal dissociation of $H_2$ on a home-made hot hydrogen atom beam source. A molecular $H_2$ beam is passed through a hot W filament held at 1900K. The pristine graphene substrate is placed 10 cm away from the filament, held at RT during atomic H deposition and subsequently cooled down to 5K, the temperature at which we carried out all STM/STS experiments presented here. $H_2$ pressure is regulated by a leak valve and fixed to $3 \cdot 10^{-7}$ torr as measured in the preparation chamber for the present experiments. The atomic H coverage was adjusted by changing the deposition times between 200-60s which corresponded to final coverages between 0.10-0.03 H atoms/nm$^2$ (or equivalently, 0.0026-0.0008ML; 1ML= 38 atoms/nm$^2$ = $3.8 \cdot 10^{15}$ atoms/cm$^2$, referred to carbon atoms in graphene layers).

After the H deposition the graphene surface presents several nearly identical point defects surrounded by threefold ($\sqrt{3} \times \sqrt{3}$) patterns. The comparison of our atomically resolved STM images of these defects with our calculations, see Fig. S1, shows that these defects correspond to single H atoms chemisorbed on graphene.

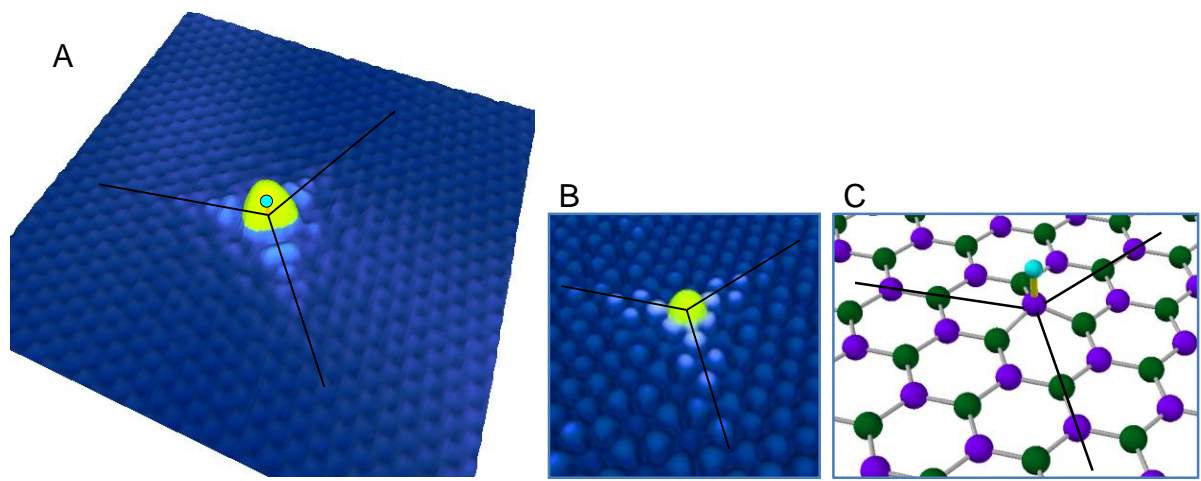

**Figure S1**. A) STM topography of a single H atom chemisorbed on graphene (0.2V, 0.1nA, 7x7nm$^2$). B) DFT simulated STM image. C) Schematic diagram showing the corresponding H adsorption site (blue ball) on graphene (purple and green balls). The three lines, at 120º each, outlined in each panel correspond to the threefold ($\sqrt{3}\times\sqrt{3}$) patterns generated on graphene by the H atom. All simulated images are calculated at the same energy as the corresponding experimental one

As control experiments, to ensure our assignment of the nature of the new threefold bright features as H atoms on graphene, we repeated exactly the same preparation procedure but either with the W filament off (i.e. H$_2$ pressure = 3·10$^{-7}$ torr; time =10 min, Temperature W filament =RT –the same holds for T$_{filament}$ < 1500K-) or without H$_2$ gas (i.e. H$_2$ pressure = 0 torr; time =10 min, temperature W filament = 1900K -we went up to T$_{filament}$ >2000K-). In both cases graphene samples looked identical to the pristine ones and no traces of H could be found on the substrates (no additional threefold bright features were detected), confirming our identification of H atoms.

The experimental data here reported were acquired at 5K by using a home-made low temperature scanning tunneling microscope (LT-STM) in ultra-high vacuum (UHV) conditions (*12*). Conductance *dI/dV* spectra were taken both using a lock-in technique, with an ac voltage (frequency: 830 Hz, amplitude: 1-2 mV rms) added to the dc sample bias, and by numerical derivative. The data were acquired and processed using the WSxM software (*36*).

**1b Identification of the atomic hydrogen chemisorption site.**

The exact atomic location of H atoms with respect to the graphene lattice can be inferred from STM measurements, as it is reflected in the complex R3 scattering patterns originated from them (*25,26,38,39*). Single H atoms chemisorbed on graphene exhibit a 3-fold scattering symmetry with the presence of three "arms" at 120º each. Thanks to this triangular symmetry, the exact identification of the H adsorption site becomes quite simple; it is just given by the intersection of the three lines which goes along the three "arms" at 120º each generated by each H atom, see Fig. S1 and S2.

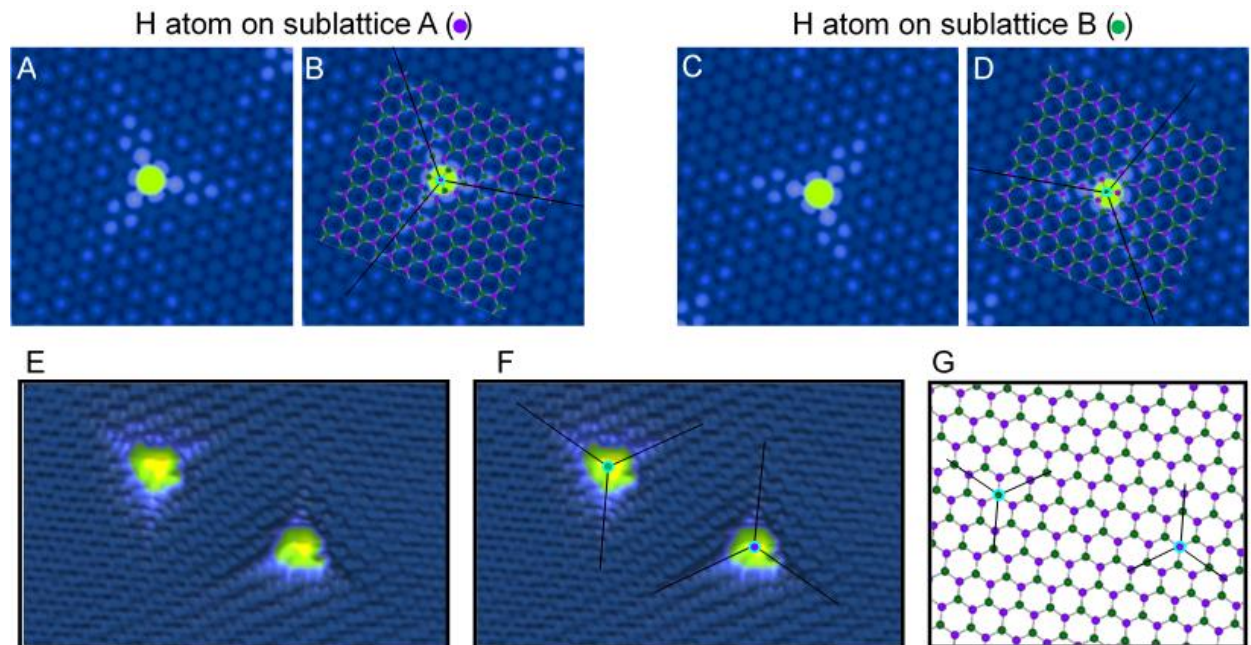

n this way, when several H atoms are present (at a distance > 0.5nm), the determination of their relative position with respect to the graphene sub-lattice, key for the present work, becomes trivial: H atoms chemisorbed on different sublattices show the three arms pointing in opposite directions, see Figs. S2 and S13.

**Figure S2**. A) Simulated STM image of a hydrogen atom chemisorbed on graphene. B) Same calculated image as A) with a schematic diagram of the graphene lattice superimposed. Three black lines are also outlined along the three "arms" at 120º each generated by each H atom to illustrate that their intersection identify the H adsorption site. C, D) same as A, B to show that a H atom in the complementary graphene sublattice has the three "arms" pointing in the opposite direction. E) Experimental STM image with two H atoms adsorbed on different sublattices (8.8x5.5 nm$^2$, $V_{bias}$ =0.4V, $I_t$=0.03nA). F) Same image as E where three black lines are outlined along the three "arms" at 120º each generated by each H atom identify the H adsorption site. G) Schematic diagram showing the corresponding H adsorption sites (blue rings) on graphene (purple and green balls) for the two H atoms shown in E, F (note that the actual experimental H-H distance is much larger).

The situation turns out a bit different for the shortest H-H distances (<0.5nm). While for H atoms on the same graphene sublattice (AA pairs) the three arms are always clearly observed independently of H-H distance, see Fig. S3A, for H atoms on different sublattices (AB pairs), at distances below 0.5nm, those legs are more difficult to be observed (although still possible in some high resolution images, not shown here). In those cases the exact H adsorption site can be inferred by direct comparison with DFT calculated images see Fig. S3 and Fig. 3 of the main manuscript.

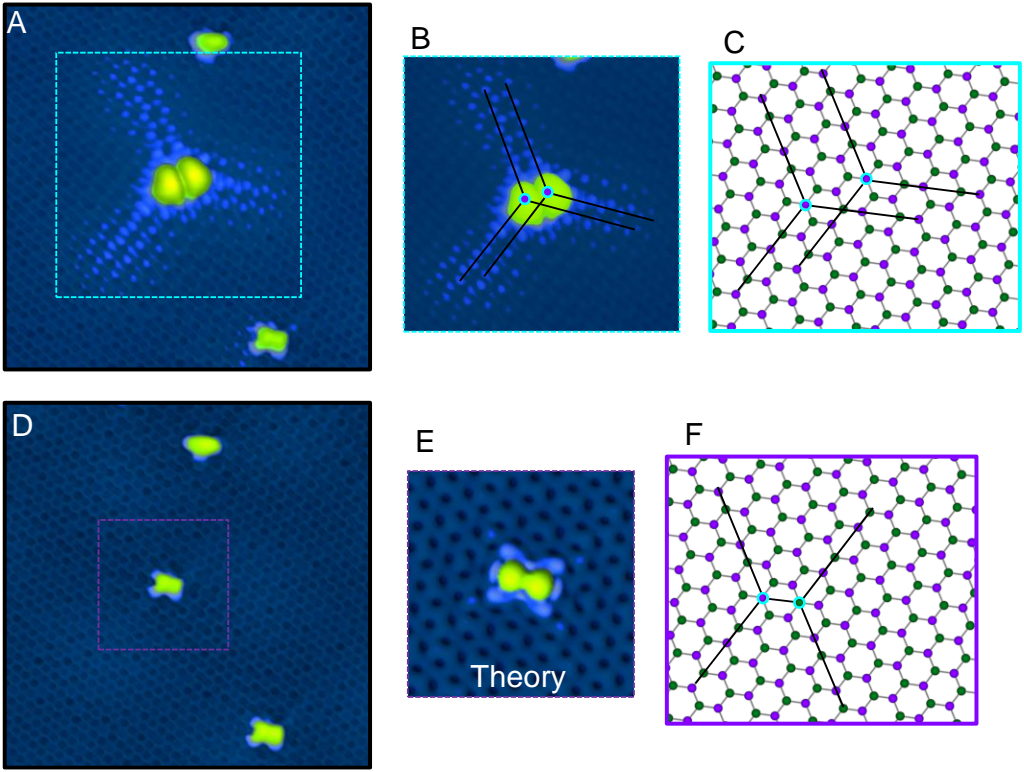

**Figure S3**. A) STM image showing different pairs of H atoms at very short H-H distances. An AA pair (outlined by the blue square) can be seen in the center of the image. B) Zoom in of the region outlined by the blue square in (A). The three "arms" at 120º each generated by both H atoms are clearly visible and are used to identify both H adsorption sites. C) Schematic diagram showing the corresponding H adsorption sites (blue rings) on graphene (purple and green balls) for the AA pair shown in (B). D) Same region as in (A) after laterally moving one H atom to form an AB dimer. E) Calculated STM image of the AB pair outlined in (D). F) Schematic diagram showing the corresponding H adsorption sites (blue rings) on graphene (purple and green balls) for the AB pair outlined in (D). Images were acquired at: $V_{bias}$=0.2V, It=0.1 nA.

For practical purposes, it is interesting to notice that for H-H distances <1.5nm, there is a very efficient way of identifying a H-H dimer as an AA or AB pair. When adsorbed on the same graphene sublattice (AA pairs) the apparent height measured by STM, for the same tunneling conditions, is ~0.5-1 Å larger than for dimers with atoms on different sublattices (AB pairs) (this criterion applies for bias voltages $V_{bias}$ ~100-400 mV). The main reason for this apparent height difference stems in the existence of a large local DOS (spin split) peak in the vicinity of $E_F$ for the case of the AA dimer which is absent for the non-magnetic AB dimers, see Fig. 4 (E, F) (one has to keep in mind that STM topography reflects the LDOS integrated from $E_F$ to $V_b$).

## 2 Influence of the underlying graphene layers.

In our work, to ensure the charge neutrality of the surface graphene layer, a "thick" (5-7 graphene layers) multilayer graphene sample was grown. It is important then to address the question about the influence of the underlying graphene layers with respect to H induced magnetism. In brief, our experiments, supported by theoretical calculations, show a small influence of the underlying graphene layers which does not affect to the main results presented here (existence of a fully polarized spin-split state, atomically modulated spin texture of the magnetic state, interaction between H atoms, how to manipulate them…).

As it is well documented in the literature, see refs (*22-24*), for large rotation angles (>10°) between the two upmost graphene layers, the surface graphene layer is electronically decoupled and can essentially be considered as a free standing graphene layer. Our experiments show that in this case there is only some residual influence of the underlying graphene layer in the LDOS of isolated H atoms, which is reflected in small variations of the relative amplitude between the empty and occupied peaks depending on the actual H atom we are measuring (most likely due to a different C-C stacking for the C atom on top of which the H is chemisorbed. For small rotation angles (<10°) between the two upmost graphene layers, the influence of the underlying graphene layer is a bit more perceptible, since the difference in the C-C stacking sequence can even modify slightly the peak-splitting, see Fig. S4C.

Those results are supported by our DFT calculations, which also show that the underlying graphene layers have little influence with respect to H induced magnetism. As previously reported by some of us (*40*) DFT calculations show that, as in the case of isolated graphene, the adsorption of H on bilayer graphene or graphite also induces a magnetic moment of $1\mu_B$, mainly, on the host monolayer. In addition, we have calculated the DOS for H atoms chemisorbed on C graphene atoms presenting different C-C stackings, in particular for H on top of $\alpha$ and $\beta$ C atoms on BL graphene and for 13° twisted bilayer graphene. As shown in Figure S4E, all calculated spectra show the spin polarization of the state independently of the stacking sequence, which only affects to the actual value of the energy spin splitting, see Figure S4F.

Our finding indicates the validity of our results to multi-layer graphene structures (i.e. independently of the stacking sequence) as long as the graphene surface remains neutral.

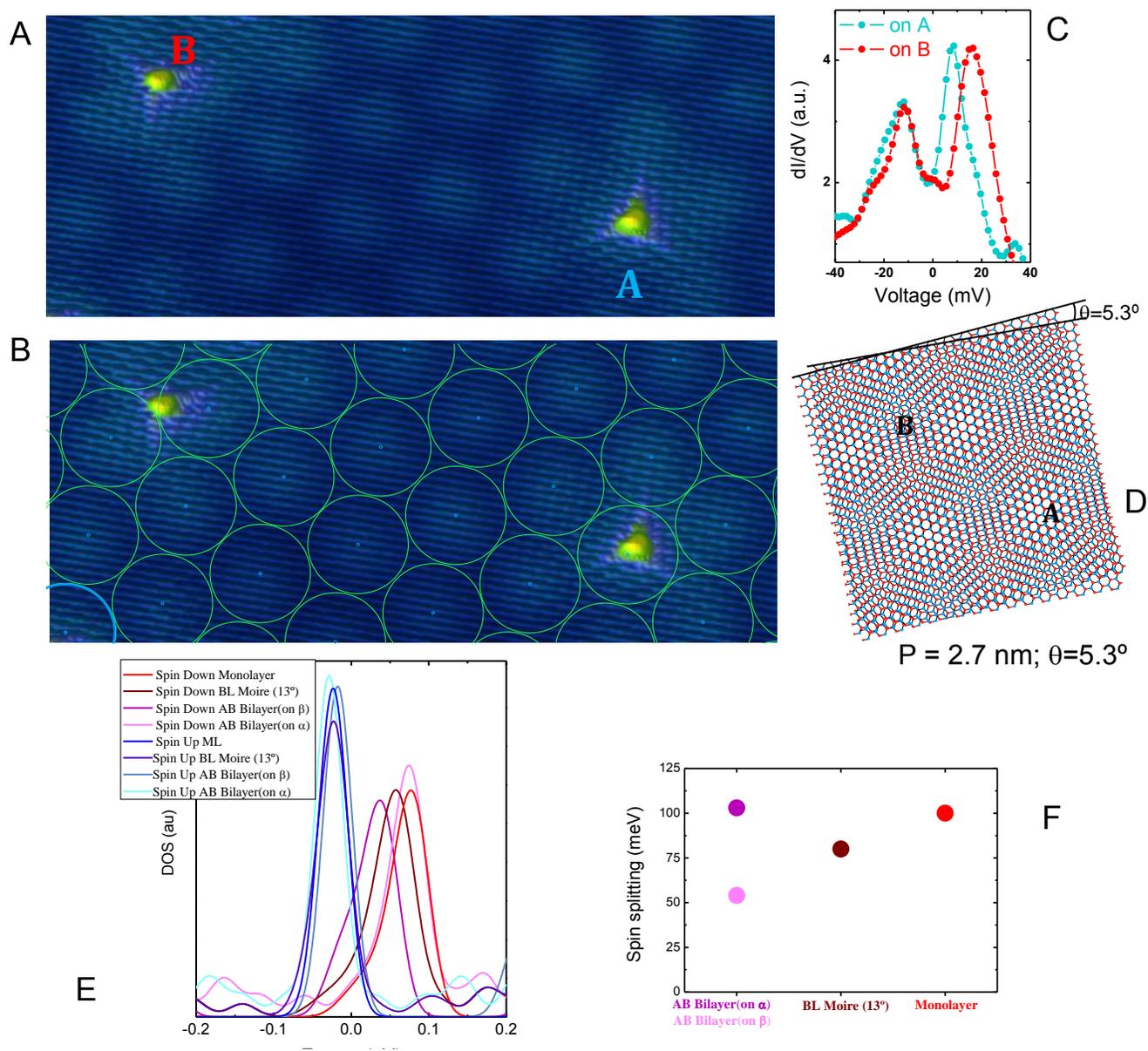

**Figure S4**. A) STM topography of a graphene moiré pattern arising from a θ = 5.3° rotation angle between the two upmost graphene layers (0.2V, 50 pA, 19x8nm$^2$). Two single H atoms chemisorbed on the bright (H atom A) and dark (H atom B) part of the moiré, which correspond to AA and AB stacking regions respectively see ref (*41*) and panel (D), can be observed. B) Same image as (A) with a schematic moiré (2.7nm periodicity, θ = 5.3°) outlined to better appreciate H adsorption site inside the moiré. C) *dI/dV* spectra on each single H atom, showing the slightly energy difference in the splitting of the polarized peak. The spectra were acquired consecutively at a nominal junction impedance of 4 gigaohm (-200 mV, 50 pA). D) Illustration of a moiré pattern arising from a θ = 5.3° rotation angle; letters indicate that H atom A is on an AA stacking region and H atom B is on an AB region. E,F) DFT simulated DOS (E) and energy spin splittings (F) for H atoms chemisorbed on graphene with different stackings.

## 3. Influence of electronic graphene doping on the spin-split state.
### 3a. Anderson Impurity model.

Here, we illustrate the influence of doping on localized magnetic moments in terms of the Anderson Impurity model, ref (*28*). This simple model contains the main ingredients required for an intuitive understanding of the charge tuning of magnetism in graphene after H adsorption. The model considers an impurity atom (described by a single electronic state "d" in the simplest case) in a host material (a metal with a featureless density of states). Anderson identified three important parameters which define the conditions for the existence or absence of localized magnetic states on this impurity: the energy position of the impurity level $E_d$ with respect to the metal Fermi level $E_F$, the energy width of this state $\Delta$ (which results from its coupling with the host material), and the intra-atomic Coulomb repulsion $U$ in the impurity state. Anderson introduced a 2D-phase diagram "to trace out the transition curve from magnetic to non-magnetic behavior", see Fig. S5 adapted from ref (*28*), with the two axes accounting for $\Delta$, and $E_F$-$E_d$ (both normalized by the effective Coulomb interaction $U$). A magnetic moments only develops when $U$ is significantly larger than $\Delta$ ($\pi\Delta/U<1$). In this case, the behavior depends on the position of $E_F$ with respect to $E_d$ and $E_d+U$: the magnetic moment, and thus the splitting, is suppressed when $E_d+U<E_F$ or $E_d>E_F$ whatever the value of $\Delta$. For a given $\Delta$ value (in our system $\Delta$ is fixed by the coupling of the H induced impurity states with the host graphene substrate), the transition from a magnetic state to a non-magnetic one can be realized via electronic doping: this would correspond to moving along a vertical line at a fixed $\pi\Delta/U$ value in the original phase diagram proposed by Anderson, see vertical dashed line in Fig. S5.

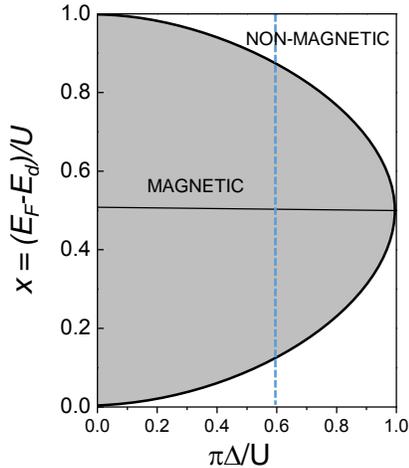

**Figure S5**. Regions of magnetic (grey shadow) and nonmagnetic (white) behaviour. Vertical dashed line outlines the magnetic-nonmagnetic transition due to electronic doping for a fix $\pi\Delta/U = 0.6$ value, see text and Fig. S6. Figure adapted from Fig. 4 of ref (*28*).

To illustrate this behavior, we plot in Figure S6 the (local) density of states for the two spin components of the impurity resonance (panels S6(a) to (e)) together with the splitting of the two spin components (central panel S6(f)) as a function of ($E_F$-$E_d$). The energy of the peaks for majority (spin ↑) and minority (spin ↓) components of the resonance are given by $E_{\uparrow(\downarrow)}= E_d+Un_{\downarrow(\uparrow)}$, where $n_{\uparrow(\downarrow)}$ are the occupation of the majority (minority) states (their splitting being $U(n_\uparrow-n_\downarrow)$), ref (28). The origin of the energy is set at the Fermi level for the plots in panels S6(a) to (e), and a value of $\pi\Delta/U =0.6$ is chosen to mimic our *dI/dV* spectra measured for H monomers on neutral graphene. The maximum splitting of the majority and minority spin components of the impurity resonance (and accordingly the maximum spin imbalance $n_\uparrow$-$n_\downarrow$) is obtained for x=0.5 (Fig. S6(a)), when the

Fermi level is located halfway between the two ↑ and ↓ peaks. When the Fermi level is shifted upwards (x>0.5) or downwards (x<0.5) with respect to $E_d$, the splitting first decreases (Fig. S6(b) and (c)) and eventually disappears (i. e. the local magnetic moment vanishes) for large or small enough x values (Fig. S6(d) and (e)).

In other words, starting from the "optimum" situation x=0.5 (maximum splitting or magnetic moment), doping the host material with electrons or holes will initially reduce the splitting (and moment) until it vanishes for sufficient strong doping. This closely resembles the behavior we observe for the H induced state on graphene (Fig. 1 of the main paper): the (spin) splitting is large for neutral graphene (when the Fermi level is located halfway between the peaks dus to the ↑ and ↓ components) and vanishes for the (strongly) doped material (notice that in graphene the energy position of the impurity levels follows the Dirac point position $E_D$, ref (29), and thus the peak(s) remains close to the Dirac point of graphene for any doping, see Fig. 1F of the main paper and SOM 3c).

The results on the influence of doping on the spin splitting obtained from the Anderson model (Fig. S6(f)) also resembles those of more realistic DFT calculations presented in the next section (Fig. S7). We thus believe that this simple and well established model provides a theoretical framework which allows an intuitive description of our results.

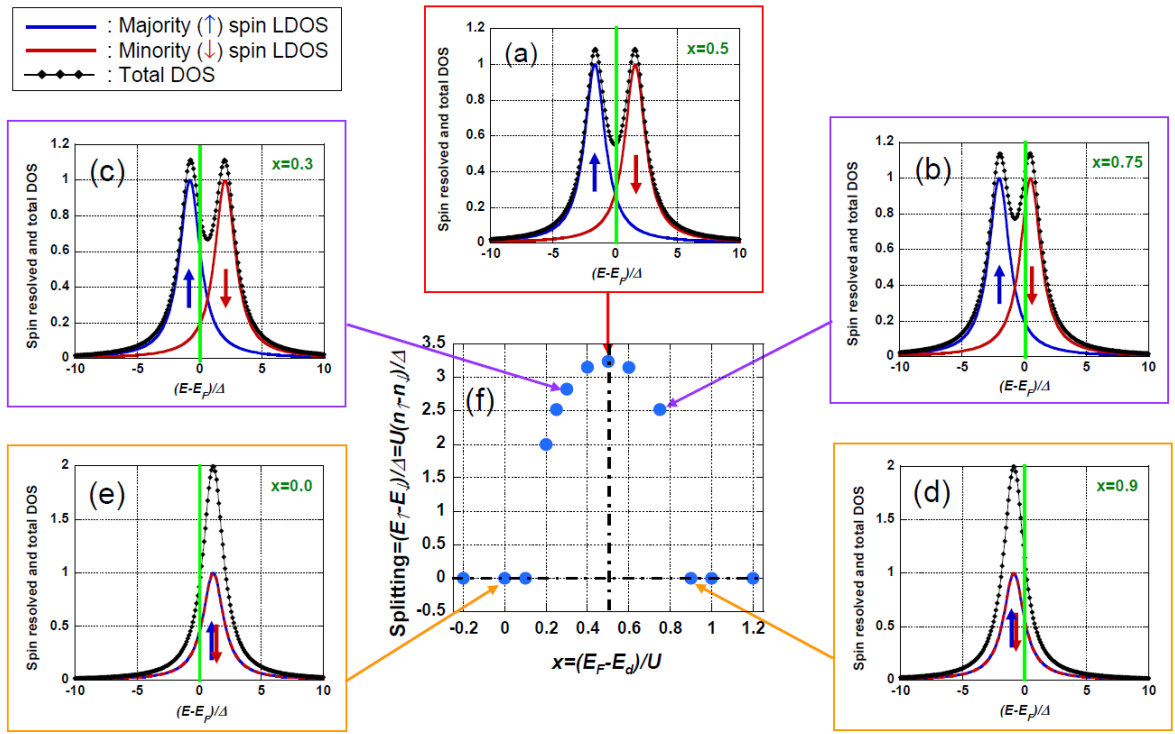

**Figure S6.** Evolution of the peak splitting with doping according to the Anderson impurity model, ref (28), with $\pi\Delta/U=0.6$ (this value is chosen to mimic our $dI/dV$ spectra measured for H monomers on neutral graphene.). $x=(E_F-E_d)/U$ describes the doping of the host material. x= 0.5, corresponds to the "optimum" case for which states $E_\uparrow$ and $E_\downarrow$ are symmetrically placed about $E_F$. For large electron or hole doping of the host material, the splitting of the states ($E_\uparrow$ and $E_\downarrow$) vanishes, corresponding to a transition to a non-magnetic state. The spin-resolved and total LDOS on the impurity is plotted in (a) to (e) for selected x values. The splitting between the majority and minority components (in units of $\Delta$), as a function of doping x, is shown in (f).

## 3b. DFT calculations.

We have also analyzed the evolution of the peak splitting with graphene-doping using more realistic DFT calculations, see sections Sect. 7-8 for details of the methodology. DFT calculations corroborate the results of the Anderson impurity model, showing that the splitting of the DOS polarized peak and the associated magnetic moment gradually decreases as we increase the electron doping of the graphene layer. When we add one electron (or hole) per super-cell -and thus per H atom-, only one occupied peak is observed in the DOS with the system becoming non-magnetic, see Fig. S7. Further electron (hole) doping just shifts the position of the DOS single peak to more negative (positive) energy values, following the change in $E_D$ position. This implies that the first extra-electron (hole) available in the system is used to fill-up (empty) the unpaired electron originated upon H chemisorption and the additional ones for overall electron (hole) doping. If the split-state had a different non-magnetic (i.e. "one electron") origin, as in the case of the non-magnetic AB dimer shown in Fig. S8 (see also Sect. 5), we should observe a rigid variation of the peak positions with doping and both of them should still be observed even when the Dirac point is far away from $E_F$ (see Fig. S8). Our results are consistent with the case of $sp^3$ defects in graphene, where it was shown that the magnetism could be controlled by doping (*15*).

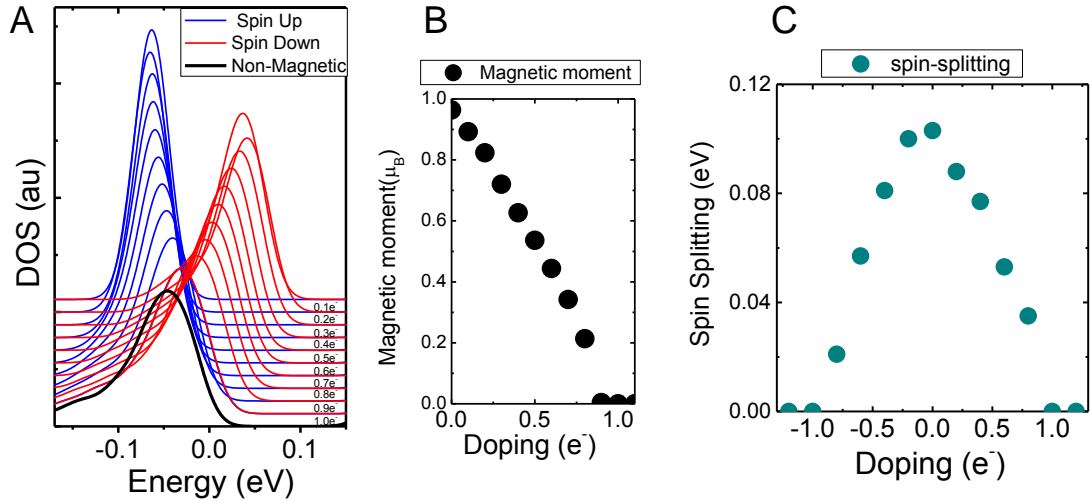

**Figure S7**. (A) Calculated DOS for an H atom on graphene for different electronic dopings between 0 and 1.0 electrons per super-cell. (B) Calculated magnetic moment induced by the H atom as a function of electron-doping (C) Calculated spin-splitting induced by the H atom as a function of electron and hole doping.

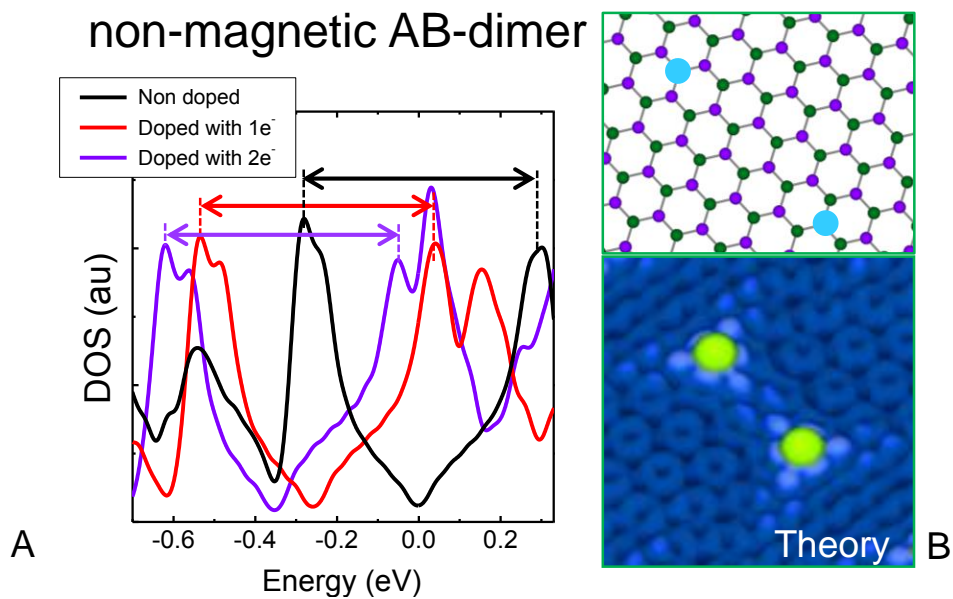

**Figure S8**. (A) Calculated DOS for a non-magnetic AB dimer on graphene for different electronic dopings of 0, 1 and 2 electrons per super-cell. A rigid shift of the two peaks in the DOS associated to the "bonding antibonding" states can be appreciated. (B) Calculated STM image and the corresponding schematic diagram showing the H adsorption sites (blue balls) on graphene (purple and green balls) for the AB dimer calculated in (A).

## 3c. STS experiments on doped graphene

As explained previously in the main manuscript and sections Sect. 3A-B, the transition from a magnetic state to a non-magnetic one can be realized by tuning the energy position of the impurity level with respect to $E_F$ via graphene-doping, which can be used as an independent experimental proof of the magnetic nature of the H-induced graphene split-state observed in neutral graphene. Therefore, we also investigated the spectroscopic signature of single H atoms on electron and hole doped graphene layers. Due to a small charge transfer from the substrate, the first graphene layers grown on SiC(000-1) are known to be slightly electron doped, graphene becoming neutral from ~5-6 layers thickness (*42,43*). Fig. S9A, shows 3 single H atoms adsorbed on a 3-layers thick graphene island on SiC(000-1). Far enough from the H atoms our STS data show a featureless V shape with its minimum, marking the position of $E_D$, at -0.14eV confirming that the graphene sheet is electron doped, see inset of Fig S9B (we have obtained the same doping level by the analysis of the quasiparticle interferences (QPIs) generated in the vicinity of atomic-size impurities such as the H adatoms (*44-46*)). STS spectra consecutively measured with the same tip on top of single H atoms show the appearance of a single occupied sharp peak, whose maximum is essentially located at the Dirac point position.

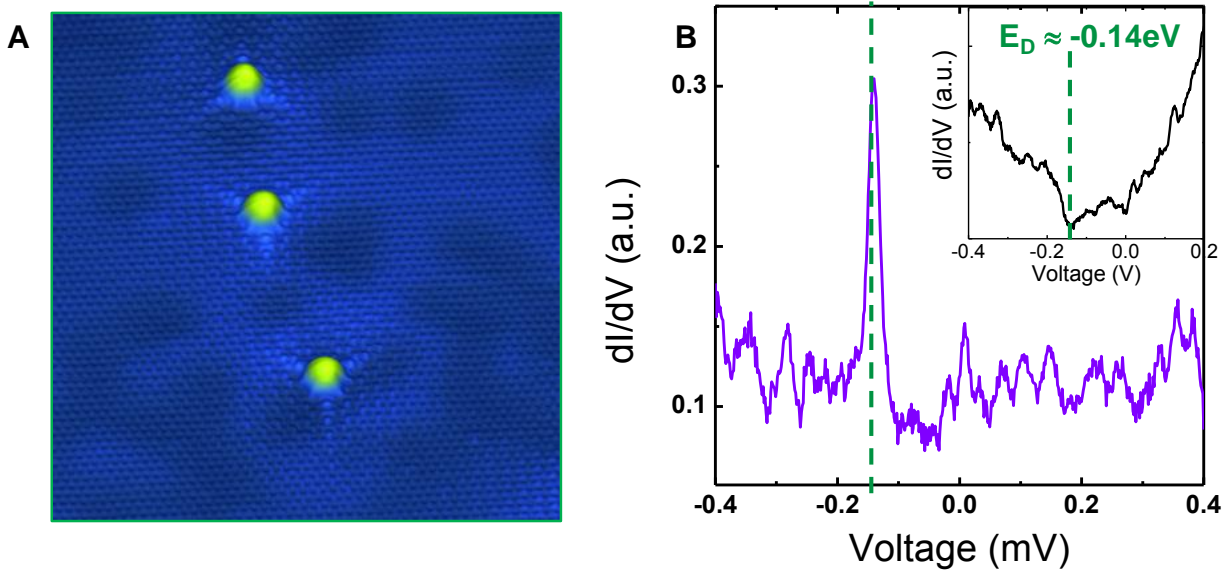

**Figure S9**. H atoms on electron-doped graphene. (A) STM topography showing three single H atom chemisorbed in 3-layers thick graphene on SiC(000-1) (-0.4V, 0.04nA, 20x20nm$^2$). (B) *dI/dV* spectra measured on single H atoms, purple curve, and clean graphene, inset. As shown in the inset, the graphene layer is electron doped with $E_D$ at -0.14eV. *dI/dV* spectra on H atoms show a single occupied peak at ~$E_D$ position.

As shown by our STS data, graphene hole-doping was achieved after exposing our samples to many cycles of H dosing and further annealing. Our STS data on hole-doped graphene (far enough from any H atom) showed a featureless V shape with its minimum, marking the position of $E_D$, at positive energies. $E_D$ values between +0.02eV and +0.1eV were typically achieved. Our *dI/dV* spectra consecutively measured on H atoms chemisorbed on those hole-doped graphene layers, showed the appearance of a single occupied sharp peak at the Dirac point position, see Fig. S10.

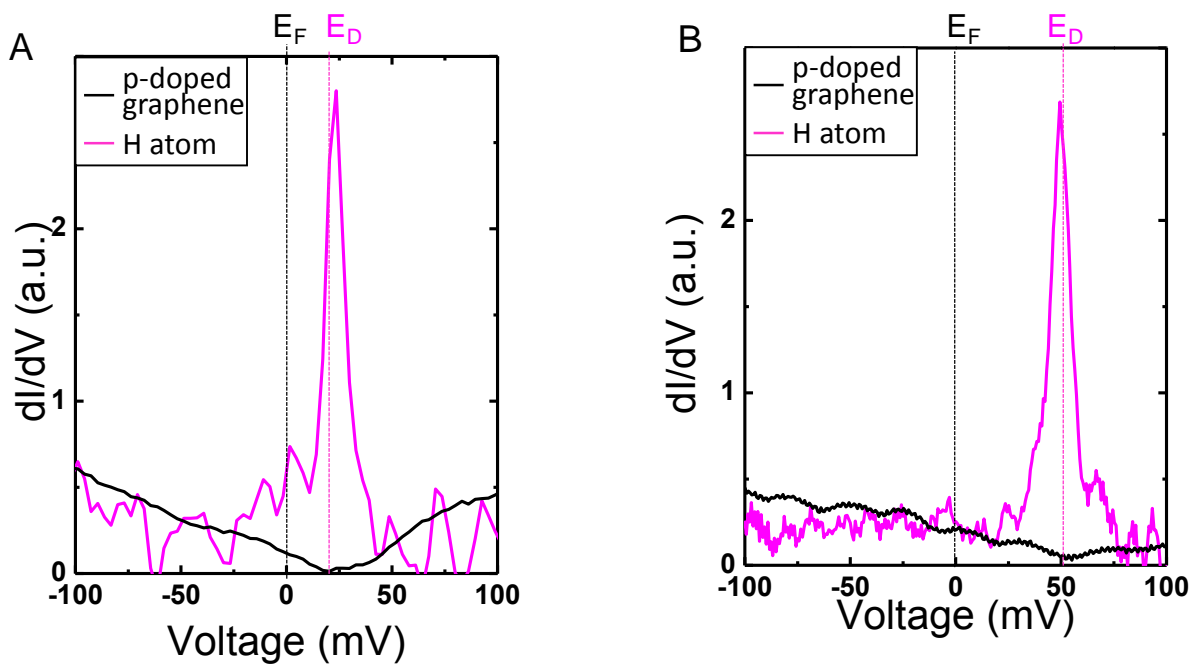

**Figure S10**. H atoms on hole-doped graphene. (A, B) *dI/dV* spectra measured on single H atoms (pink curve) and clean graphene (black curve) on graphene layers with different hole-doping levels. The graphene layer are hole-doped with $E_D$ at +0.025eV (A) and $E_D$ at +0.050eV (A). For both doping levels, *dI/dV* spectra on H atoms show a single occupied peak at ~$E_D$ position.

## 4. Spatial extension of the spin-polarized electronic state.

Fig. S11A shows a map of the *dI/dV* vs energy, measured along the 6-nm line across the H atom drawn in Fig. S11B. The green (purple) balls mark the position of the C atoms along the profile in the same (opposite) sublattice to that of the H atom. The states presents strong atomic-scale modulations of the peak intensities. To better appreciate this critical sublattice dependence, in Fig. S11C we have plotted the *dI/dV* curves acquired on two neighboring C atoms (m and n) belonging to different sublattices, which we have extracted from the vertical dashed lines outlined in the LDOS($x,E$) map of Fig. S11A. The spin-polarized state only emerges in the C atom belonging to opposite (purple) sublattice showing no significant weight on the C atom from the same (green) sublattice. As shown in Fig. S11D, the calculated projected density of states (PDOS) for those two particular C atoms is in very good agreement with our experimental findings.

The experimental *dI/dV* data obtained here is formed by both spin-up and spin-down contributions. However, since in this system the spin-split state is fully polarized, this is enough to get information about the relative magnetization of each graphene site. Our DFT calculations show that the height of the occupied LDOS peak measured on a carbon atom is very closely correlated with its local magnetic moment, see Figs. S11D-F. Thus, by measuring the spatial evolution of the *dI/dV* occupied peak height, we can experimentally map the spatial distribution of the local magnetic moment induced by the atomic H chemisorption. This is further plotted in Fig. S11G presenting the magnetic moment distribution along the dashed line drawn in Fig. S11B, which is extracted from the *dI/dV*(x,-10meV) horizontal dotted line outlined on the conductance map of Fig. S11A. Our data show that the magnetic moment is essentially induced on the carbon atoms belonging to the graphene sublattice opposite to H adsorption.

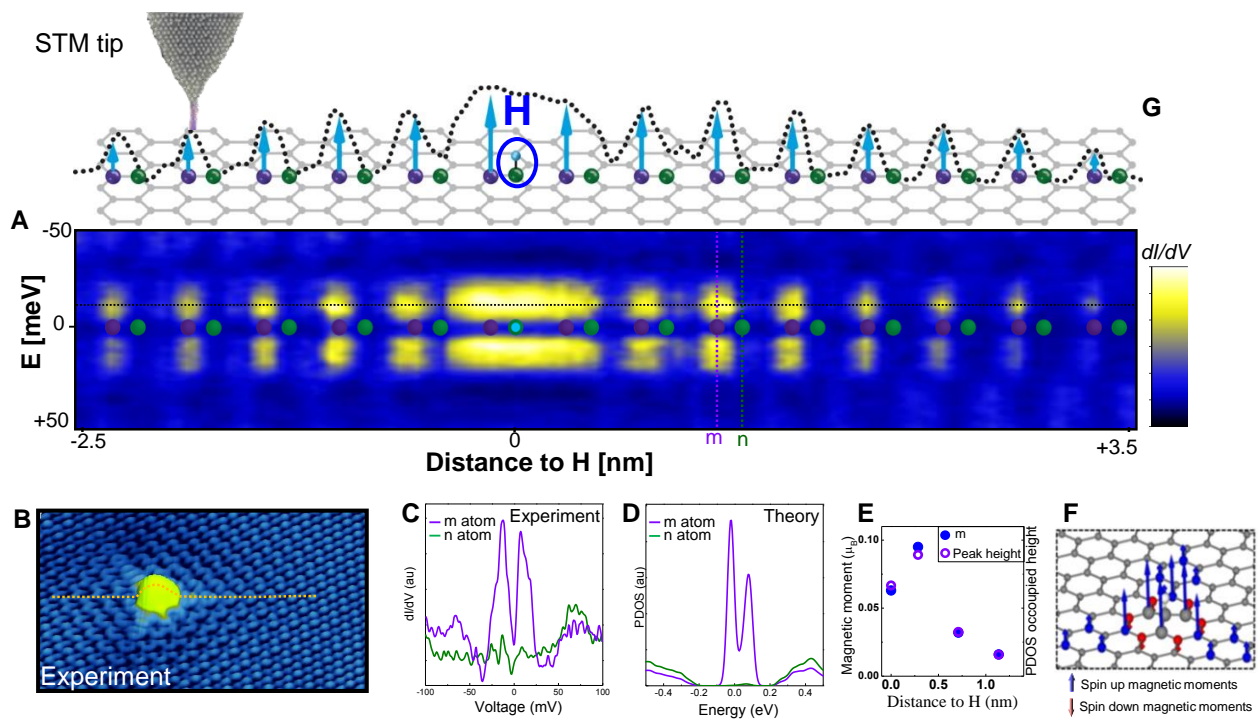

**Figure S11**. (A) Spatially resolved conductance map - *dI/dV(x, E)* - along the dashed line drawn in (B). Green/purple balls indicate the positions of C atoms along the profile. (B) STM topography of a single H atom on graphene (0.2V, 0.1nA, 7x5nm$^2$). (C) *dI/dV* curves measured on the m and n carbon atoms outlined in (A). (D) PDOS calculated on those m and n carbon atoms. (E) Comparison between local magnetic moment and occupied PDOS peak height calculated on different C atoms. (F) Calculated magnetic moments induced by H chemisorption. (G) Schematic diagram of the graphene structure along the profile drawn in (B), green/purple balls indicating the positions of C atoms belonging to the same/opposite sublattice with respect to H chemisorption. The dotted line shows the evolution of the occupied peak height and the arrows the relative magnetic moment contribution of each C atom, see text.

## 5. H-H magnetic coupling at large distances.

Our STS experiments show a marked difference on the LDOS associated to both type of dimers: AA dimers present a fully polarized state close to $E_F$ which is absent in AB dimers. Our STS data also show that this sublattice dependent magnetic coupling persists for H dimers separated by many atomic sites. To illustrate this point, we show in Fig. S12A two H atoms in AB configuration separated by 1.15nm, as inferred from our calculated images, see Fig. 12B. Again, our STS spectra showed no traces of the polarized state confirming that the AB coupling inhibits magnetism even at such large distances, see Figs. 12C, D with experimental and theoretical data respectively.

We have also computed the exchange energy. The exchange energy for 2H atoms in AA sites at distance 14.96 Å is: **$E_{ex}$ [(2H in AA with spin 1) – (2H in AA forced to spin 0)] = −0.0353** eV

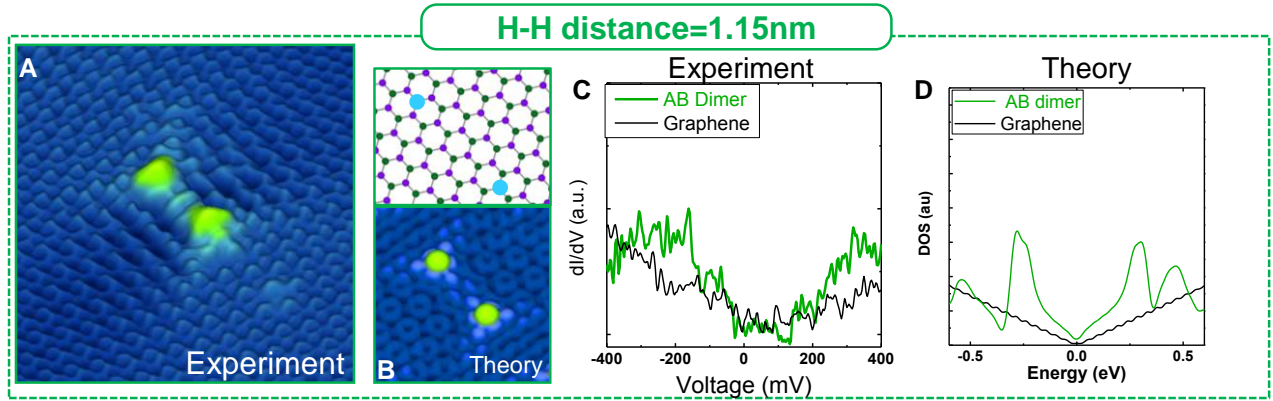

**Figure S12.** H-H coupling at large distances. (A) STM topography of an AB dimer separated by 1.15 nm (-0.2V, 0.05nA, 10x10nm$^2$). (B) Calculated STM image, within the Tersoff-Hamann approximation, and the corresponding schematic diagram showing the H adsorption sites (blue balls) on graphene (purple and green balls) for the AB dimer shown in (A). (C, D) Experimental (C) and calculated (D) DOS generated by the AB dimer and clean graphene.

## 6. Atomic H manipulation

Choosing the appropriate tunneling parameters atomic H can be removed, laterally moved, and even deposited on graphene surfaces with atomic precision. The role played by the STM tip is to selectively modify the binding energy landscape of H atoms to produce the required H manipulation, see Sect. 10 for DFT calculations on the H extraction.

To selectively remove H atoms from the graphene sample we approached the STM tip towards the sample. This can be done by continuously increasing, under feedback control, the setpoint tunneling current on top of the selected H atom until it is desorbed, or by switching the feedback off and slightly decreasing the tip-sample distance on top of it. It is also possible to completely remove all H atoms from a graphene region by imaging it at high currents. The precise tunneling values for the manipulation depend on each specific tip apex, but for the same tip apex those values are very reproducible. As a rough guide, for removing H atoms it is usually enough to approach the STM tip by 1-2 Å.

The deposition of H atoms is done by applying negative sample voltages pulses. In order to deposit H atoms, we first need to pick them up from the graphene surface, so the tip can act as an H reservoir. Then by applying negative sample voltages pulses the H atoms are deposited on the selected graphene region under the tip position (see Fig. S13). Again, the voltage threshold for H deposition might vary from tip to tip, but values of around -5 V are usually enough for H deposition.

Finally, small positive sample voltages enable the lateral manipulation of H atoms.

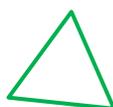 7 H atoms on "B"

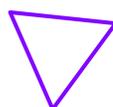 7 H atoms on "A"

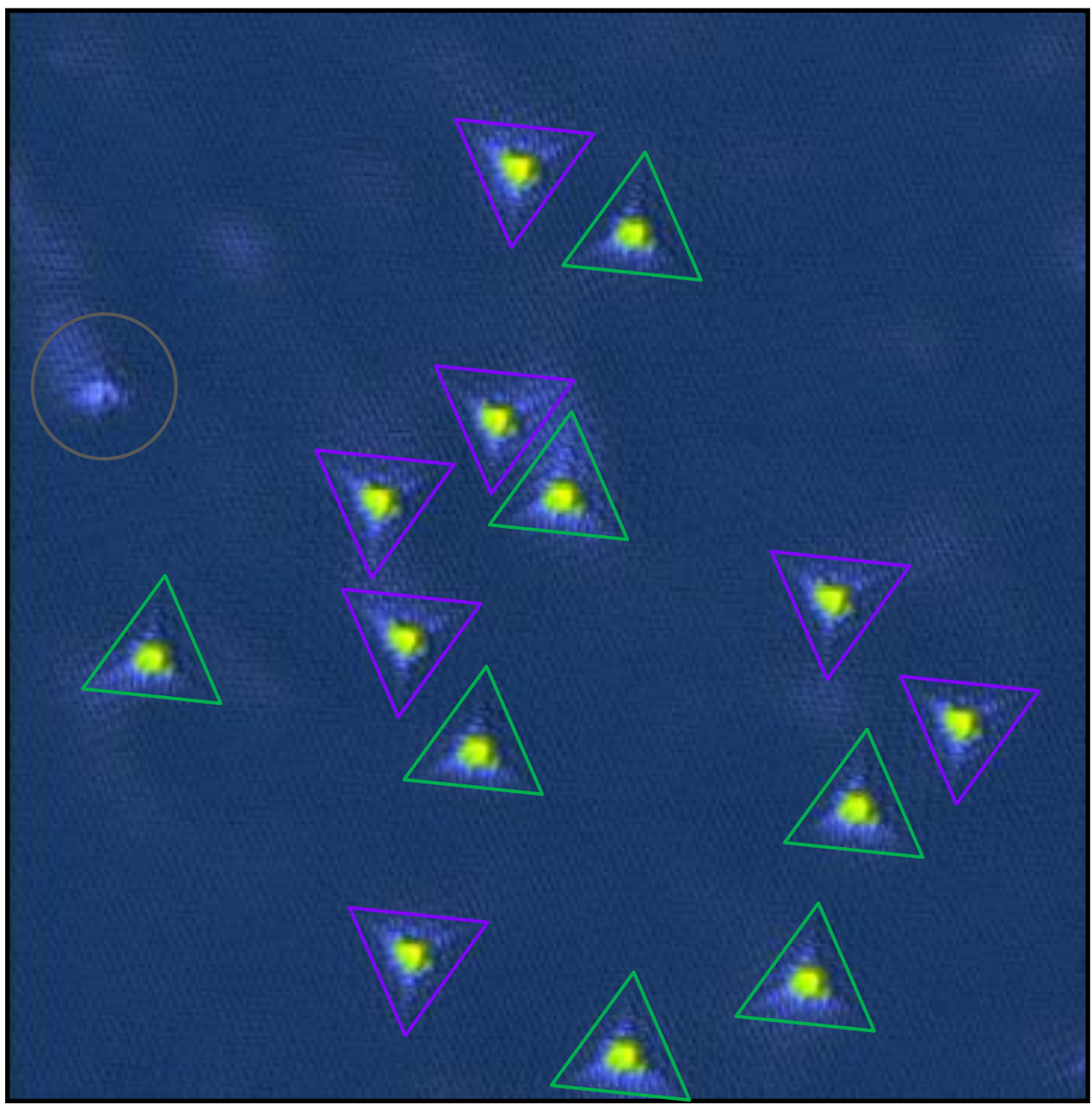

**Figure S13.** Raw STM data showing the same image as in Fig. 4J of the main manuscript represented at larger size. It shows 14 H atoms that we selectively deposited on a clean graphene region to reach a final configuration with 7 H atoms chemisorbed on each graphene sublattice as outlined by green and purple triangles. The gray circle on the left hand-side outlines a sub-surface defect. Image size 28x28 nm$^2$, $V_{bias}$ =0.4V, $I_t$=0.03nA.

## 7. Theoretical methodology H on graphene

The hexagonal lattice of graphene is formed by carbon atoms which are connected by means of $sp^2$ hybridized orbitals. This electronic structure results in three σ orbitals contained in the graphene plane forming 120° angles between each other and one π orbital in the axis perpendicular to the graphene.

For the Hydrogen atom to be chemisorbed, the π bonds need to be broken to form a new σ bond. The physical mechanism which allows such a transition is the local deformation of the crystallographic structure by a carbon atom moving out of the graphene plane, thus transforming the $sp^2$ hybridization to a local $sp^3$ hybridization (*17,18*).

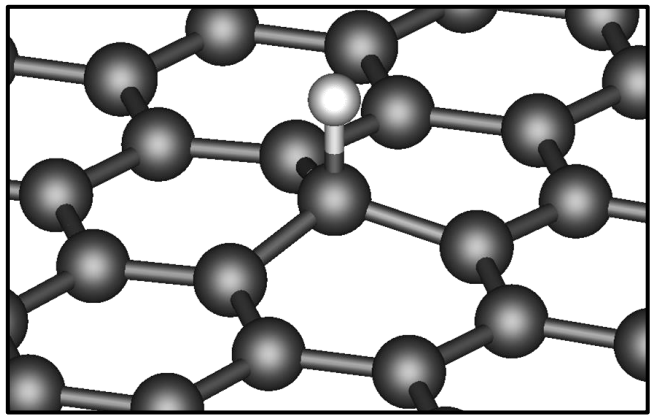

**Figure S14**. DFT atomic model of the hydrogen atom chemisorption on graphene.

In order to study the geometrical and electronic structure of the different defects in graphene we use the first principles density functional (*47, 48*) SIESTA code (*49, 50*) which uses localized orbitals as basis functions (*51*). We use a double ζ basis set, non-local norm conserving pseudopotentials and for the exchange correlation functional we use the local density approximation (LDA). The results have being checked with generalized gradient approximation (GGA) (*52*) calculations. The calculations are performed with stringent criteria in the electronic structure convergence (down to $10^{-5}$ in the density matrix), 2D Brillouin zone sampling (up to 1600 k-points), real space grid (energy cut-off of 400 Ryd) and equilibrium geometry (residual forces lower than $3 \times 10^{-2}$ eV/Å). Due to the rapid variation of the density of states at the Fermi level, we used a polynomial smearing method (*53*).

To study defects we use the super cell approximations in the way that we end up with an interaction between defects in the repeated unit cell. To minimize this interaction we use unit cells of different size and, in addition, we use "skewed" unit cells in a way that the lattice vector do not coincide with graphene symmetry directions therefore interactions along the zig-zag and armchair chains of atoms are minimized, see Fig. S15.

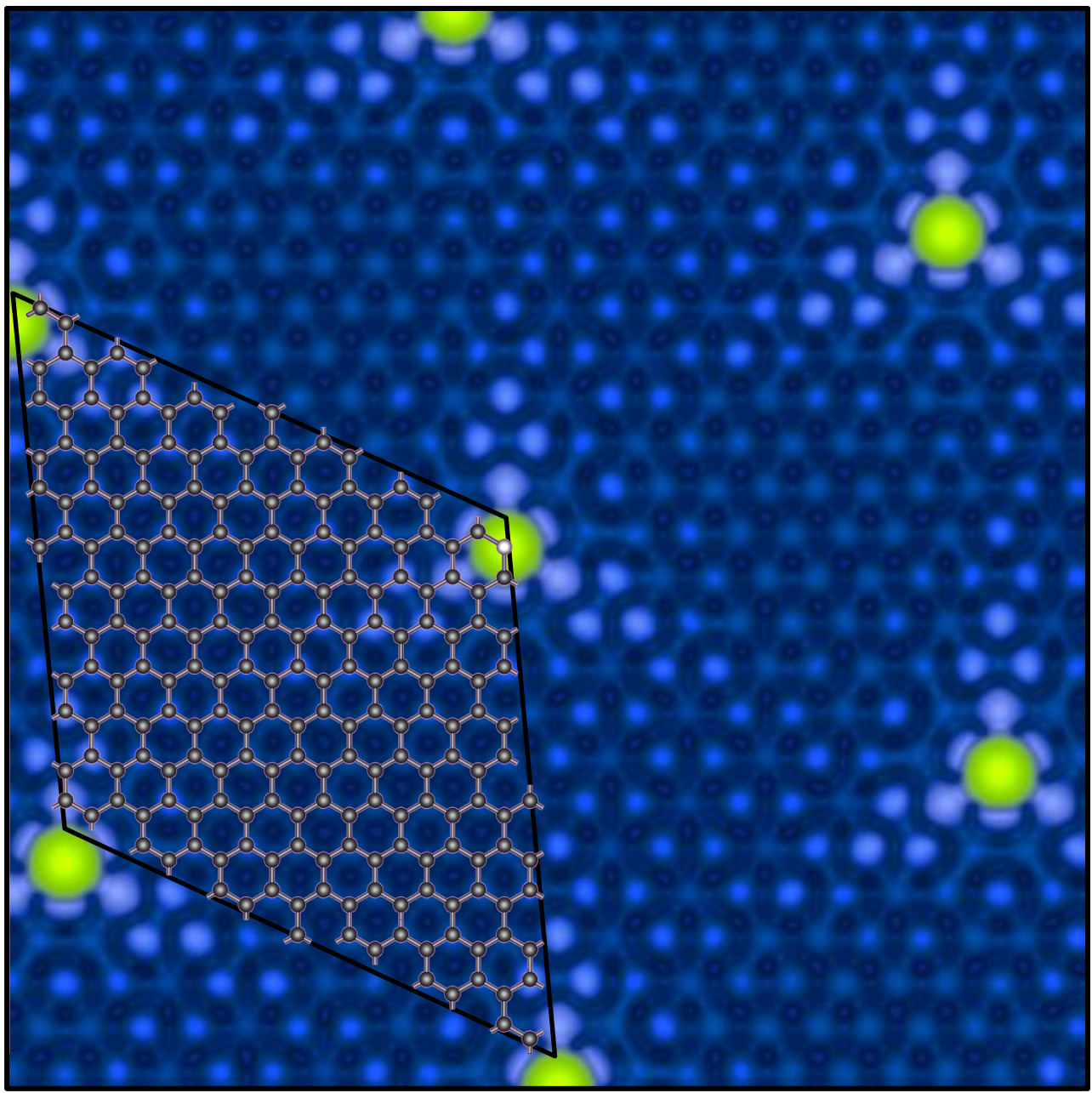

**Figure S15.** Simulated STM image, within the Tersoff-Hamann approximation, of a hydrogen atom in a 218 atoms graphene unit cell. The atomic model is superimposed to outline the (5,7) "skewed" character of the unit cell.

## 8. Band structure and density of states calculated by DFT.

In Fig. S16 we show the results of a non spin-resolved calculation of one hydrogen atom in a 218 atoms graphene unit cell. We immediately realize the existence of an essentially half occupied extremely narrow peak at the Dirac point. The charge transfer between hydrogen and graphene being of the order of a few hundredths of electron.

Results of the corresponding spin resolved calculations are shown in Fig. S17. We immediately notice the spin-up spin-down splitting of the hydrogen- induced state at the Dirac point. In the fully relaxed geometry, not only the carbon atom, which binds directly to the hydrogen one, but the three surrounding ones move up to favor the process. The Carbon-Hydrogen distance becomes 1.20 Å, with the central carbon moved 0.35 Å upwards and the three surrounding around 0.08 Å. The $sp^3$ hybridization advantages the elongation of the hydrogenated graphene, and thus the lift of the Carbon atoms, see Fig. S14. The calculated adsorption energy of the Hydrogen on graphene being around 1 eV.

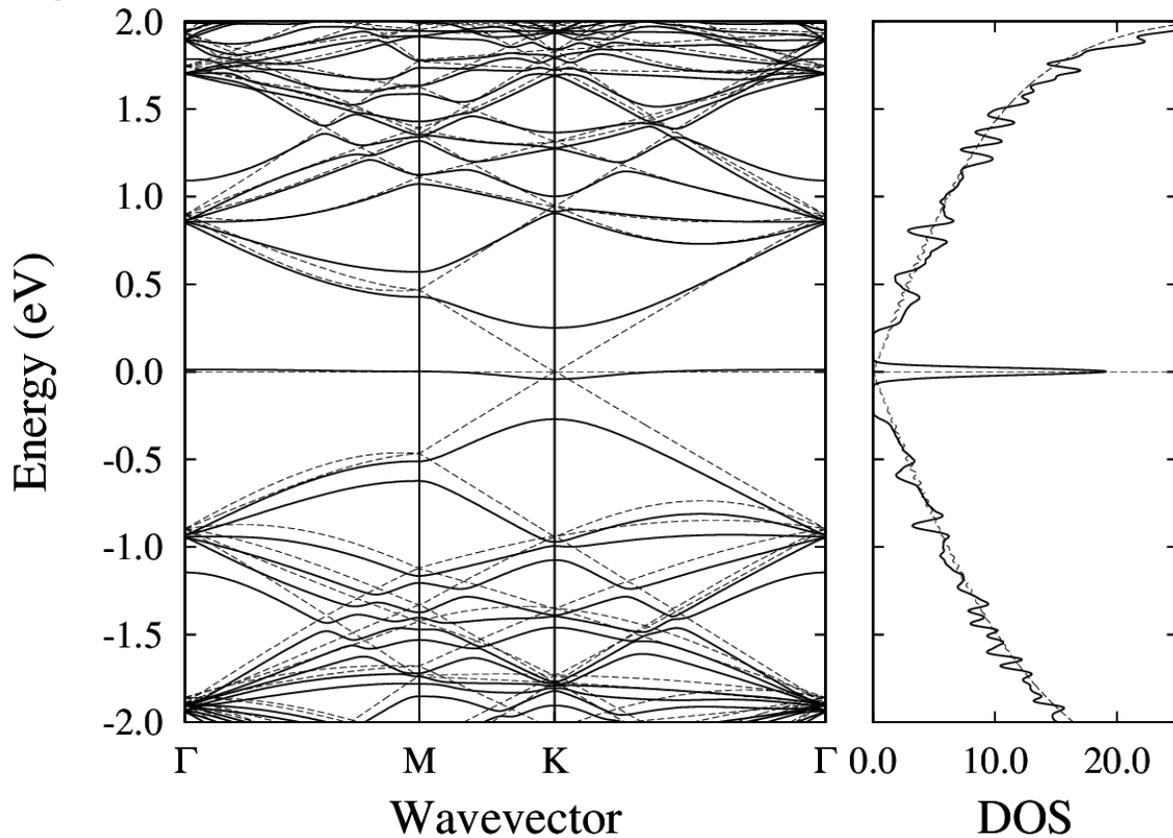

**Figure S16**. Non spin-resolved band structure (left panel) and density of states (right panel) of a hydrogen atom in a 218 atoms graphene unit cell (Fig. S8). The horizontal dotted line indicates the Fermi energy. A small (0.025 eV) gaussian broadening is included in the density of states for presentation purposes. The dash line indicates the defect free graphene results.

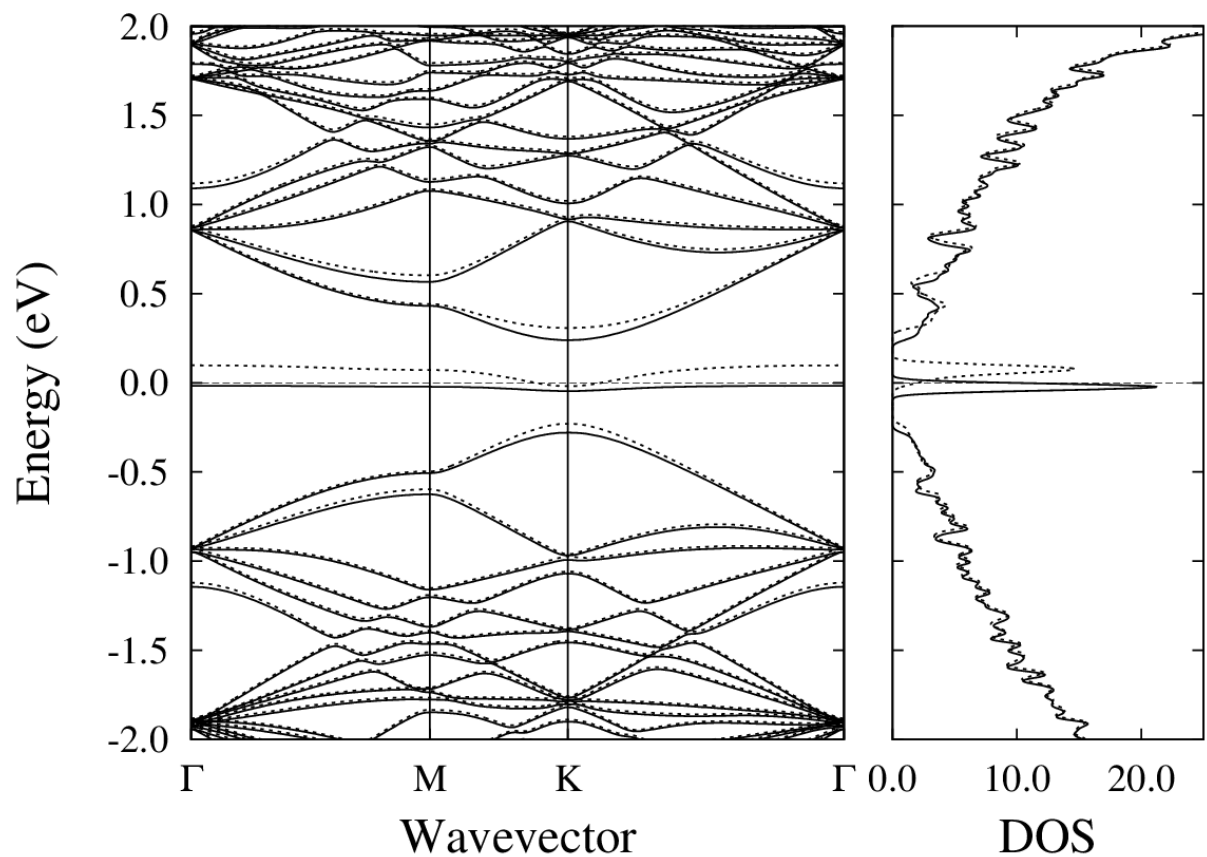

**Figure S17**. Spin-resolved band structure (left panel) and density of states (right panel) of a hydrogen atom in a 218 atoms graphene unit cell. Solid and broken lines indicate spin up and spin down states respectively. The horizontal dotted line indicates the Fermi energy. A small (0.025 eV) gaussian broadening is included in the density of states for presentation purposes.

## 9. Energy spin-splitting dependence with unit cell size.

We have studied how the spin-up spin-down peaks splitting depends on the size of the super-cell used in the calculation and therefore on the localization of the hydrogen induced state at the Fermi level as well as on the parasitic interaction between adjacent cells. Results of the calculations are shown in Fig. S18. We notice that, as expected, ref (*4*), the splitting decreases with the size of the super-cell. As the size of the super-cell increases the state is more delocalized, the effective electron-electron interaction U decreases and the energy splitting of the state is smaller (see the sketch of Figure 1-A of the main text). We have performed the calculations with super-cell sizes up to a 26x26- (1352+1 atoms). The results indicate that the energy splitting extrapolates to a finite non-zero value.

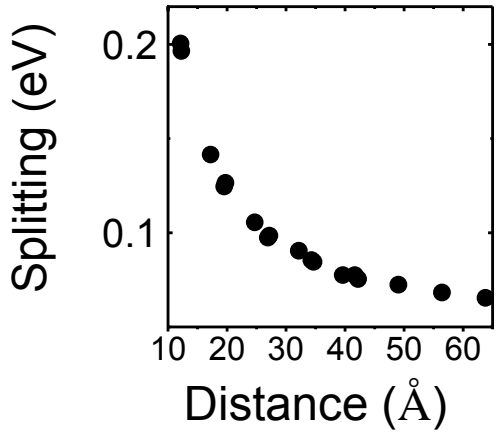

**Figure S18.** Super-cell calculation of the energy splitting between spin up and spin down peaks and its dependence on the distance between hydrogen atoms in adjacent cells.

## 10. Pt tip manipulation of H

To understand why a Pt STM tip can perform the manipulations shown in Fig. 4 of the main manuscript, we have carried out DFT calculations for a model system as the one shown in Fig. S19 where a pyramidal Pt tip approaches a H atom adsorbed on graphene. The calculated adsorption energy of a H atom on the Pt tip apex is around 2.8 eV. Because of the difference between the adsorption energy for H on graphene (around 1 eV) and that on the Pt tip, it can be inferred that H prefers to adsorb on the Pt tip, the only obstacle being the desorption barrier. This barrier can be suppressed by approaching progressively the tip to the deposited H. Figure S19 shows the evolution of the binding energy curve of the H atom in between the tip and graphene. When the tip gets closer both adsorption energy minima merge at some point. When retracting the tip, the desorption barrier builds up again, but now with the H adsorbed on the Pt tip (see also movie S1). The blue and red arrows indicate the spin density on the atoms as in the main text. The Pt tip apex also develops a magnetic moment, but we have not depicted it for clarity.

The manipulation of H with the Pt tip has been modelled through the DFT+ Green's function methodology as implemented in ANT.G (*54-56*). The DFT part in this code is performed by Gaussian (*57*). Only the electrons from the $5s^2$, $5p^6$, $5d^9$ and $6s^1$ shells of Pt were treated explicitly, while the shells below these were replaced by Effective Core Potentials (ECP), as is the case of the LanL2DZ basis set in the Gaussian code. In the case of H, the same type of basis set reproduces the complete electronic structure without the need of ECP potentials. Finally, the bonding Carbon atom was treated with the complete basis set LanL2DZ describing the six electrons, while the other carbon atoms were described by the minimum basis set CRENBS in which only the $2s^2$ and $2p^2$ shells are treated explicitly, with the others replaced by an ECP potential. The functional used was BLYP, which typically leads to proper results for covalent and metallic bonds in both organic and metallic elements. The functional used was BLYP (*58*), which typically leads to proper results for covalent and metallic bonds in both organic and metallic elements, complemented with dispersion forces through the GD3 Grimme implementation (*59*).

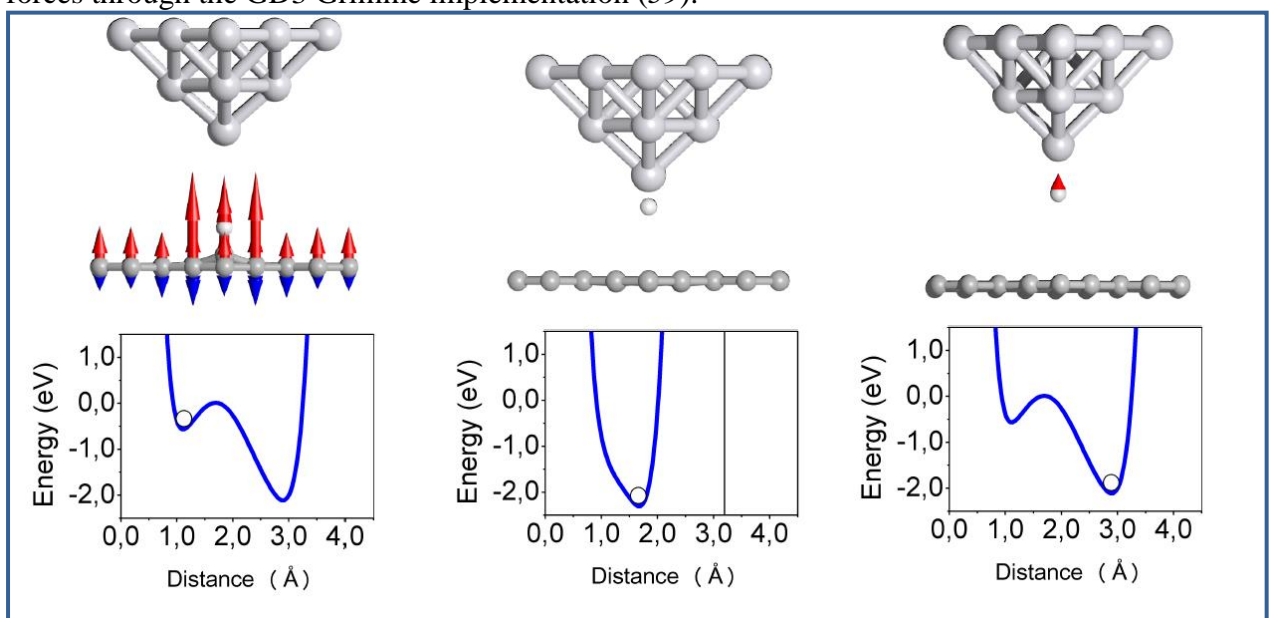

**Figure S19**. DFT calculation of the evolution of the binding energy of the H atom in between the tip and graphene as a function of tip-sample distance. See movie S1 for the complete sequence.

## 11. dI/dV spectrum on the AA dimer of figure 4E in the vicinity of $E_F$

Experimental *dI/dV* spectra shown in Fig. 4 were measured, with identical tunneling conditions, immediately before and after the H manipulation. Particular care was taken to ensure that the STM tip DOS remained unmodified during the manipulation procedure, the main goal being to show the emergence or disappearance of a spin-split state on the graphene layer after the H manipulation. Figure S20 shows a *dI/dV* spectrum measured on the same AA dimer as in Fig. 4E, with higher resolution in the vicinity of $E_F$, revealing a double occupied peak (at negative bias) as in Figure 3E.

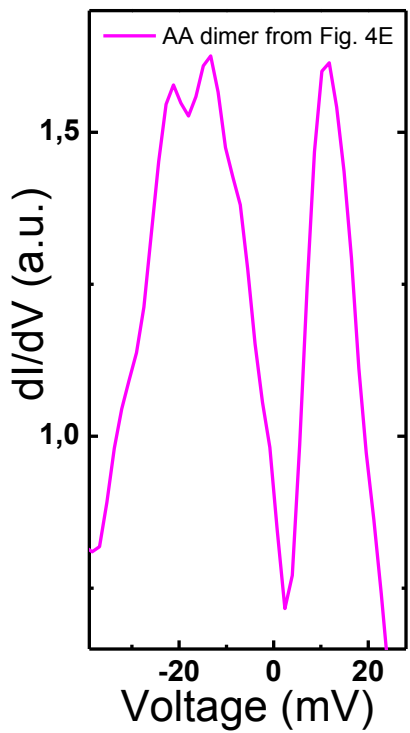

**Figure S20**. Experimental *dI/dV* spectra for the AA dimer shown in Fig. 4E of the main manuscript. The spectra was acquired at a nominal junction impedance of 8 GΩ (-400 mV, 50 pA) at 5 K.

**References:**

1. K. S. Novoselov *et al.*, *Science* 306, 666 (2004).
2. P. O. Lehtinen, A. S. Foster, Y. C. Ma, A. V. Krasheninnikov, R. M. Nieminen, *Physical Review Letters* 93, 187202 (2004).
3. V. M. Pereira, F. Guinea, J. dos Santos, N. M. R. Peres, A. H. C. Neto, *Physical Review Letters* 96, 036801 (2006).
4. O. V. Yazyev, L. Helm, *Physical Review B* 75, 125408 (2007).
5. O. V. Yazyev, *Reports on Progress in Physics* 73, 056501 (2010).
6. E. J. G. Santos, A. Ayuela, D. Sanchez-Portal, *New Journal of Physics* 14, 043022 (2012).
7. E. H. Lieb, *Physical Review Letters* 62, 1201 (1989).
8. J. Fernandez-Rossier, J. J. Palacios, *Physical Review Letters* 99, 177204 (2007).
9. J. J. Palacios, J. Fernández-Rossier, L. Brey, *Physical Review B* 77, 195428 (2008).
10. M. Fujita, K. Wakabayashi, K. Nakada, K. Kusakabe, *Journal of the Physical Society of Japan* 65, 1920 (1996).
11. C. G. Tao *et al.*, *Nature Physics* 7, 616 (2011).
12. M. M. Ugeda, I. Brihuega, F. Guinea, J. M. Gomez-Rodriguez, *Physical Review Letters* 104, 096804 (2010).
13. R. R. Nair *et al.*, *Nature Physics* 8, 199 (2012).
14. K. M. McCreary, A. G. Swartz, W. Han, J. Fabian, R. K. Kawakami, *Physical Review Letters* 109, 186604 (2012).
15. R. R. Nair *et al.*, *Nature Communications* 4, 2010 (2013).
16. J. Balakrishnan, G. K. W. Koon, M. Jaiswal, A. H. C. Neto, B. Ozyilmaz, *Nature Physics* 9, 284 (2013).
17. L. Jeloaica, V. Sidis, *Chemical Physics Letters* 300, 157 (1999).
18. X. W. Sha, B. Jackson, *Surface Science* 496, 318 (2002).
19. D. W. Boukhvalov, M. I. Katsnelson, A. I. Lichtenstein, *Physical Review B* 77, 035427 (2008).
20. D. Soriano *et al.*, *Physical Review Letters* 107, 016602 (2011).
21. F. Varchon, P. Mallet, L. Magaud, J. Y. Veuillen, *Physical Review B* 77, 165415 (2008).
22. M. Sprinkle *et al.*, *Physical Review Letters* 103, 226803 (2009).
23. G. H. Li *et al.*, *Nature Physics* 6, 109 (2010).
24. G. T. de Laissardiere, D. Mayou, L. Magaud, *Nano Letters* 10, 804 (2010).
25. H. A. Mizes, J. S. Foster, *Science* 244, 559 (1989).
26. P. Ruffieux, O. Groning, P. Schwaller, L. Schlapbach, P. Groning, *Physical Review Letters* 84, 4910 (2000).
27. Materials and methods are available as supporting material on *Science* Online
28. P. W. Anderson, *Physical Review* 124, 41 (1961).
29. B. Uchoa, V. N. Kotov, N. M. R. Peres, A. H. C. Neto, *Physical Review Letters* 101, 026805 (2008).
30. L. Hornekaer *et al.*, *Physical Review Letters* 97, 186102 (2006).
31. Z. Sljivancanin *et al.*, *Journal of Chemical Physics* 131, 084706 (2009).
32. R. Balog *et al.*, *Journal of the American Chemical Society* **131**, 8744 (2009).
33. D. M. Eigler, E. K. Schweizer, *Nature* 344, 524 (1990).
34. T. C. Shen *et al.*, *Science* 268, 1590 (1995).
35. A. A. Khajetoorians *et al.*, *Science* 339, 55 (2013).
36. I. Horcas *et al.*, *Review of Scientific Instruments* 78, 013705 (2007).

37. F. Hiebel *et al*., *J. Phys. D: Appl. Phys*. **45**, 154003 (2012)
38. K. F. Kelly, D. Sarkar, G. D. Hale, S. J. Oldenburg, N. J. Halas, *Science* **273**, 1371 (1996).
39. K. F. Kelly, N. J. Halas, Surface Science 416, L1085 (1998).
40. M. Moaied, J. V. Alvarez, J. J. Palacios, *Physical Review B* **90**, 115441 (2014).
41. I. Brihuega et al., *Phys. Rev. Lett*. **109**, 196802 (2012).
42. D. Sun et al., *Phys. Rev. Lett.* **104**, 136802 (2010).
43. J. Hicks, K. Shepperd, F. Wang, and E. H. Conrad, J. Phys. D: Appl. Phys., **45**, 154002 (2012).
44. G. M. Rutter, J. N. Crain, N. P. Guisinger, T. Li, P. N. First, and J. A. Stroscio, *Science* **317**, 219 (2007).
45. I. Brihuega et al., *Phys. Rev. Lett*., **101** 206802 (2008).
46. P. Mallet et al*., Phys. Rev. B* **86**, 045444 (2012).
47. P. Hohenberg and W. Kohn, *Phys. Rev*. *B* **136**, 864 (1964).
48. W. Kohn and L. J. Sham, *Phys. Rev*. **140**, 1133 (1965).
49. P. Ordejon, E. Artacho, and J. M. Soler, *Phys. Rev. B* 53, R10441 (1996).
50. J. Soler et al., *J. Phys.: Condens. Matter* **14**, 2745 (2002).
51. O. F. Sankey and D. J. Niklewski, *Phys. Rev. B* **40**, 3979 (1989).
52. J. P. Perdew and Y. Wang, *Phys. Rev. B* **45**, 13244 (1992).
53. M. Methfessel and A. T. Paxton, *Phys. Rev. B* **40**, 3616 (1989).
54. J. J. Palacios, A. J. Pérez-Jiménez, E. Louis, J.A. Vergés, *Phys. Rev. B* **64**, 115411 (2001).
55. J. J. Palacios, A. J. Pérez-Jiménez, E. Louis, E. San Fabián, J. A. Vergés, *Phys. Rev. B* **66**, 035322 (2002)
56. D. Jacob, J. J. Palacios, *J. Chem. Phys*. **134**, 044118 (2011). http://alacant.dfa.ua.es
57. M. J. Frisch et al. ,Gaussian, Inc., Wallingford CT, 2009. http://www.gaussian.com
58. B. Miehlich, A. Savin, H. Stoll, and H. Preuss, *Chem. Phys. Lett.*, **157** 200-06 (1989).
59. S. Grimme, J. Antony, S. Ehrlich and H. Krieg, *J. Chem. Phys*. **132,** 154104 (2010).

**Movie S1**

This movie shows the complete sequence for the DFT calculation of the evolution of the binding energy of the H atom in between the tip and graphene as a function of tip-sample distance.